\documentclass[useAMS,usenatbib,a4paper]{mn2e}
\usepackage{epsfig,bm}
\usepackage{amsbsy,amssymb,amsmath}
\usepackage{hyperref}
\usepackage{graphicx}
\usepackage{float}
\usepackage{subfigure}
\usepackage{natbib}
\usepackage[normalem]{ulem}

\newcommand{\be}{\begin{equation}}
\newcommand{\ee}{\end{equation}}

\newcommand{\bn}{\mathbf{\hat{n}}}
\newcommand{\bk}{\mathbf{k}}
\newcommand{\bea}{\begin{eqnarray}}
\newcommand{\eea}{\end{eqnarray}}

\def\vhel{\ifmmode{V_{{\rm HEL}}}\else{$V_{{\rm HEL}}$}\fi}
\def\vsys{\ifmmode{V_{\rm sys}}\else{$V_{\rm sys}$}\fi}
\def\kms{\ifmmode{~{\rm km\,s}^{-1}}\else{~km~s$^{-1}$}\fi}
\def\vlsr{\ifmmode{v_{\rm lsr}}\else{$v_{\rm lsr}$}\fi}
\def\ltsim{\ifmmode\stackrel{<}{_{\sim}}\else$\stackrel{<}{_{\sim}}$\fi}
\def\gtsim{\ifmmode\stackrel{>}{_{\sim}}\else$\stackrel{>}{_{\sim}}$\fi}

\title[HI intensity mapping : a single dish approach]
{HI intensity mapping : a single dish approach}
\author[R.A. Battye {\it et al}]
{R.A. Battye$^{1}$, I.W.A. Browne$^{1}$, C. Dickinson$^{1}$, G. Heron$^{1}$ , B. Maffei$^{1}$, A. Pourtsidou$^{1,2}$ \\
$^{1}$Jodrell Bank Centre for Astrophysics, School of Physics and Astronomy, University of Manchester, Oxford Road, Manchester M13 9PL, U.K.\\
$^{2}$Dipartimento di Astronomia, Universit\`a di Bologna, via Ranzani 1,
 I-40127 Bologna, Italy
}
\pagerange{\pageref{firstpage}--\pageref{lastpage}}

\begin{document}

\maketitle

\begin{abstract}
We discuss the  detection of large scale HI intensity fluctuations
using a single dish approach with the ultimate objective of measuring
the Baryonic Acoustic Oscillations (BAO) and constraining the
properties of dark energy. To characterise the signal we present 3D power spectra, 2D angular power
spectra for individual redshift slices, and also individual line-of-sight  spectra,
computed using the ${\rm S}^3$ simulated HI catalogue which is based
on the Millennium Simulation.  We consider optimal instrument design and survey strategies for a single dish observation at low and high redshift for a fixed sensitivity. For a survey corresponding to an instrument with $T_{\rm sys}=50\,{\rm K}$, 50 feed horns and 1 year of observations, we find that at low redshift  ($z\approx 0.3$), a resolution of $\sim 40\,{\rm arc min}$ and a survey of $\sim 5000\,{\rm deg}^2$ is close to optimal, whereas at higher redshift ($z\approx 0.9$)  a resolution of $\sim 10\,{\rm arcmin}$ and $\sim 500\,{\rm deg}^2$ would be necessary - something which would be difficult to achieve cheaply using a single dish. Continuum foreground
emission from the Galaxy and extragalactic radio sources are
potentially a problem. In particular, we suggest that it could be that the dominant extragalactic 
foreground comes from the clustering of very weak sources. We assess its amplitude and discuss ways by
which it might be mitigated. We then introduce our concept for a
dedicated single dish telescope designed to detect BAO at low redshifts. It involves an under-illuminated
static $\sim 40\,{\rm m}$ dish and a $\sim 60$ element receiver array
held $\sim 90\,{\rm m}$ above the under-illuminated dish. Correlation receivers will be
used with each main science beam referenced against an antenna
pointing at one of the Celestial Poles for stability and control of
systematics. We make sensitivity estimates for our proposed system and
projections for the uncertainties on the power spectrum after 1 year of
observations. We find that it is possible to measure the acoustic
scale at $z\approx 0.3$ with an accuracy $\sim 2.4\%$ and that $w$
can be measured to an accuracy of 16\%.
\end{abstract}


\section{Introduction}
Precision cosmology is now a reality. Measurements of the Cosmic
Microwave Background (CMB), Large-Scale Structure (LSS) and Type Ia
supernovae have led the way in defining the standard cosmological
model, one based on a flat Friedmann-Robertson-Walker Universe containing
cold dark matter (CDM) and a cosmological constant, $\Lambda$, with
densities relative to the present day critical density $\Omega_{\rm
m}\approx 0.27$ and $\Omega_{\Lambda}\approx 0.73$ \citep{Riess:1998,
Perlmutter:1999, Eisenstein:2001, Percival:2002, Komatsu:2011}. The
standard model has just 6 parameters, but these can been expanded to
constrain, rule out, or detect more exotic possibilities. One
particular case is an expanded dark energy sector, often
parameterised by the equation of state parameter $w=P/\rho$ for the
dark energy \citep{Copeland:2006}, or even the possibility of
modifications to gravity \citep{Clifton:2011}. Observations which can
constrain the dark sector are usually geometrical (standard rulers or
candles), or based on constraining the evolution of structure \citep{Weinberg:2012}.

The use of Baryonic Acoustic Oscillations (BAOs) to constrain $w$ is
now an established technique \citep{Eisenstein:1998,
Eisenstein:2003}. Originally applied to the Third Data Release (DR3) of
the SDSS Luminous Red Galaxy (LRG) survey \citep{Eisenstein:2005},
subsequent detections have been reported by a number of groups
analysing data from 2dF \citep{Cole:2005}, SDSS DR7
\citep{Percival:2010}, WiggleZ \citep{Blake:2011}, 6dF
\citep{Beutler:2011} and BOSS \citep{Anderson:2012}.
 All these have used redshift surveys performed in the optical
 waveband to select galaxies. The power spectrum of their three
 dimensional distribution is computed and, assuming that the power
 spectrum of the selected galaxies is only biased relative to the
 underlying dark matter distribution by some overall scale-independent
 constant, the acoustic scale can then be extracted. Present sample
 sizes used for the detection of BAOs are typically $\sim 10^{5}$
 galaxies, but there are a number of projects planned to increase this
 to $\sim 10^{7}$--$10^{8}$ using optical and near infra-red (NIR) observations which
 would lead to increased levels of statistical precision
 \citep{Abbott:2005, Hawken:2011, Laureijs:2011, Schlegel:2011}.

It is possible to perform redshift surveys in the radio waveband,
using 21cm radiation from neutral hydrogen (HI) to select galaxies
which could provide improved confidence in the results from the
optical/NIR. At present the largest surveys have only detected $\sim
10^{4}$ galaxies \citep{Meyer:2004}, limited by the low luminosity of
the 21cm line emission \citep{Furlanetto:2006}, but there are a number
of proposed instruments with large collecting area capable of
increasing this substantially in the next decade \citep{Abdalla:2005,
Duffy:2008, Abdalla:2010}. By the time the Square Kilometre Array
(SKA)\footnote{www.skatelescope.org} is online redshift surveys of
galaxies selected in the radio should have comparable statistical
strength to those made in the optical/NIR. The advantage of
doing both radio and optical surveys is that they will be subject to
completely different systematics.

An intriguing possibility is the idea of using what has become known
as 21cm intensity mapping. The fact that the 21cm line is so weak
means that it requires $\sim 1\,{\rm km}^2$ of collecting area to
detect $L_\star$ galaxies at $z\sim 1$ in an acceptable period of time
- this was the original rationale for the SKA
\citep{Wilkinson:1991}. However, instruments with apertures of $\sim 100\,{\rm m}$ size
have sufficient surface brightness sensitivity to detect HI at higher
redshift, but they will preferentially detect objects of angular size
comparable to their resolution as pointed out in
\citet{Battye:2004}. These objects will be clusters with masses
$10^{14}$--$10^{15}\,M_{\odot}$ and other large scale
structure. \citet{Peterson:2006} and \citet{LoebWyithe:2008} suggested that
the full intensity field $T(\theta,\phi,f)$ could be used to measure
the power spectrum as a function of redshift directly (as is done for
the CMB intensity fluctuations) if the smooth continuum
emission, e.g. that from the Galaxy, can be accurately subtracted and
any systematic instrumental artefacts, arising from the observational
strategy, can be calibrated to sufficient accuracy. If these obvious
difficulties can be mitigated, then this technique can be enormously
powerful: the standard approach of detecting galaxies and then averaging
them to probe large scale structure is inefficient in the sense that
to detect the galaxies, at say $5\sigma$, one throws away all but a
fraction of the detected 21cm emission. Although this is not a problem in the optical, in the radio where the signal is much weaker one cannot afford to take this hit. In order to measure the power
spectrum one is not trying to detect the actual galaxies, although
this can lead to a plethora of other legacy science, and all the detected
emission is related to the LSS structure. \citet{Chang:2008} have
shown that dark energy can be strongly constrained with a radio
telescope significantly smaller, and cheaper, than the SKA. 

Interferometric techniques naturally deal with some of the issues
related to foregrounds and instrumental systematics. In particular,
they can filter out more efficiently the relatively smooth and bright
contaminating signals from the Galaxy and local signals such as the
Sun, Moon and ground. However, interferometers can be extremely
expensive since they require electronics to perform the large number of correlations. In this paper we will present our ideas on the design of a
single dish telescope and an analysis of its performance (BINGO; {\bf
B}AO from {\bf I}ntegrated {\bf N}eutral {\bf G}as {\bf
O}bservations). Attempts have already been made to detect the
signal using presently operational, single-dish radio telescopes,
\citep{Chang:2010} and recent surveys of the local
universe \citep{Pen:2008}. It has been possible to make detections using cross
correlation with already available optical redshift surveys, but not  the
detection of the crucial autocorrelation signal needed to probe
cosmology using radio observations alone. The reasons for this are unclear, but they are likely to be
due to a combination of the observations not been taken with this
specific purpose in mind and the fact that the telescope and detector
system are not sufficiently stable. Our philosophy in this paper will be to quantify
the expected HI signal, quantify the contaminating foreground signals
and then present our design for a system which will be sufficiently
sensitive and stable for the BAO signal to be extracted.

\section{Characterisation of the signal}

In this section we will present some basic material on the 21cm line emission in order to characterize various aspects of the signal expected in a HI intensity mapping experiment.

\subsection{Mean Temperature}

One can compute the emitted brightness temperature of the 21cm line in
a velocity width $dv$ using \be T_{\rm emit}dv=\frac{\hslash
c^3}{k}\frac{3A_{21}}{16f_{\rm emit}^2}n_{\rm HI}dl, \ee where $k$ is
Boltzmann's constant, $T_{\rm emit}$ is the  HI brightness
temperature, $f_{\rm emit}\approx 1420.4\,{\rm MHz}$ is the rest frame
emission frequency for neutral hydrogen, $n_{\rm HI}$ is the number
density of neutral hydrogen atoms, $dl$ is the line of sight distance and $A_{21}$ is the
spontaneous emission coefficient of the 21cm transition. Equation (1) can be
used to deduce the standard formula which relates the observed integrated flux
from an object (typically a galaxy) with HI mass $M_{\rm HI}$
\citep{Roberts:1974} \be {M_{\rm HI}}={2.35\times
10^{5}\,M_{\odot}\over 1+z}{S_{\rm obs} dv\over {\rm Jy}\,{\rm
km}\,{\rm s}^{-1}}\left({d_{\rm L}(z)\over {\rm Mpc}}\right)^2\,, \ee
where $d_{\rm L}(z)$ is the luminosity distance.

This standard formula is for an isolated object. However, on the very largest scales
relevant to us here it is more convenient to work with the brightness
temperature. The mean observed brightness temperature  due to the average HI
density in the Universe is \be
\label{eq:Tobs}
\bar{T}_{\rm obs}(z)=\frac{\bar{T}_{\rm
emit}(z)}{1+z}=\left(\frac{\hslash c^3}{k}\frac{3A_{21}}{16f_{\rm
emit}^2M_{\rm{H}}}\right) \left(\frac{\rho_{\rm
HI}(z)}{1+z}\right)\frac{dl}{dv}\,,  \ee where $M_{\rm H}$ is the mass
of the hydrogen atom and $\rho_{\rm HI}(z)$ is the average density of
HI at redshift $z$. The first term is just a combination of fundamental
constants and measured experimental parameters, the second is a function of redshift and the final,
relatively unfamiliar, term relates the line of sight distance to the recession velocity and is given by $H_0^{-1}$ in the local Universe, where $H_0$ is the Hubble constant.

In order to compute the last term in Equation (\ref{eq:Tobs}), let us consider the comoving volume element
\be
\frac{dV}{dzd\Omega}=\frac{cr^2}{H(z)}=c\left(\frac{r}{1+z}\right)^2\frac{(1+z)^2}{H(z)}=c \ [d_{\rm A}(z)]^2\frac{(1+z)^2}{H(z)},
\ee
\be 
r(z)=c\int_0^z\,{dz^{\prime}\over H(z^{\prime})}\,,
\ee
where $d_{\rm A}(z)=r(z)/(1+z)$ is the angular diameter distance and 
. We can also write $dV=d_{\rm A}^2dl\,d\Omega$ and hence
\be
\frac{dl}{dz}=\frac{c}{H(z)}(1+z)^2.
\ee
Therefore, using $dv/c=dz/(1+z)$, we find that
\be
\frac{dl}{dv}=\frac{(1+z)^3}{H(z)}.
\ee

Hence,  (\ref{eq:Tobs}) becomes
\be
\label{eq:Tobs1}
\bar{T}_{\rm obs}(z)=\left(\frac{3A_{21}\hslash c^3}{16f_{\rm emit
}^2kM_{H}}\right)\frac{(1+z)^2\rho_{HI}(z)}{H(z)},
\ee
and substituting in values for the fundamental constants and re-writing $\rho_{\rm HI}(z)$ in terms of the density relative to the present day critical density, $\Omega_{\rm HI}(z)=8\pi G\rho_{\rm HI}(z)/(3H_{0}^2)$, gives
\be
\bar{T}_{\rm obs}(z)=44 \mu {\rm K}\left(\frac{\Omega_{\rm HI}(z)h}{2.45 \times 10^{-4}}\right) \frac{(1+z)^2}{E(z)}\,,
\label{tave}
\ee
where $h=H_0/100\,{\rm km}\,{\rm sec}^{-1}\,{\rm Mpc}^{-1}$ and $E(z)=H(z)/H_0$.

Throughout we will assume that $\Omega_{\rm HI}h=2.45\times 10^{-4}$ independent of redshift which is the value calculated from the local HI mass function measured by HIPASS~\citep{Zwaan:2005}. Both observational~\citep{Prochaska:2009} and theoretical arguments~\citep{Duffy:2011} appear to suggest that there is no more than a factor of 2 change in density of neutral hydrogen over the range of redshifts relevant to intensity mapping. In the rest of the paper we will drop the ``obs'' subscript and use $T$ to denote the observed brightness temperature.

\subsection{3D power spectrum}

\begin{figure}
\centering
\includegraphics[scale=0.3]{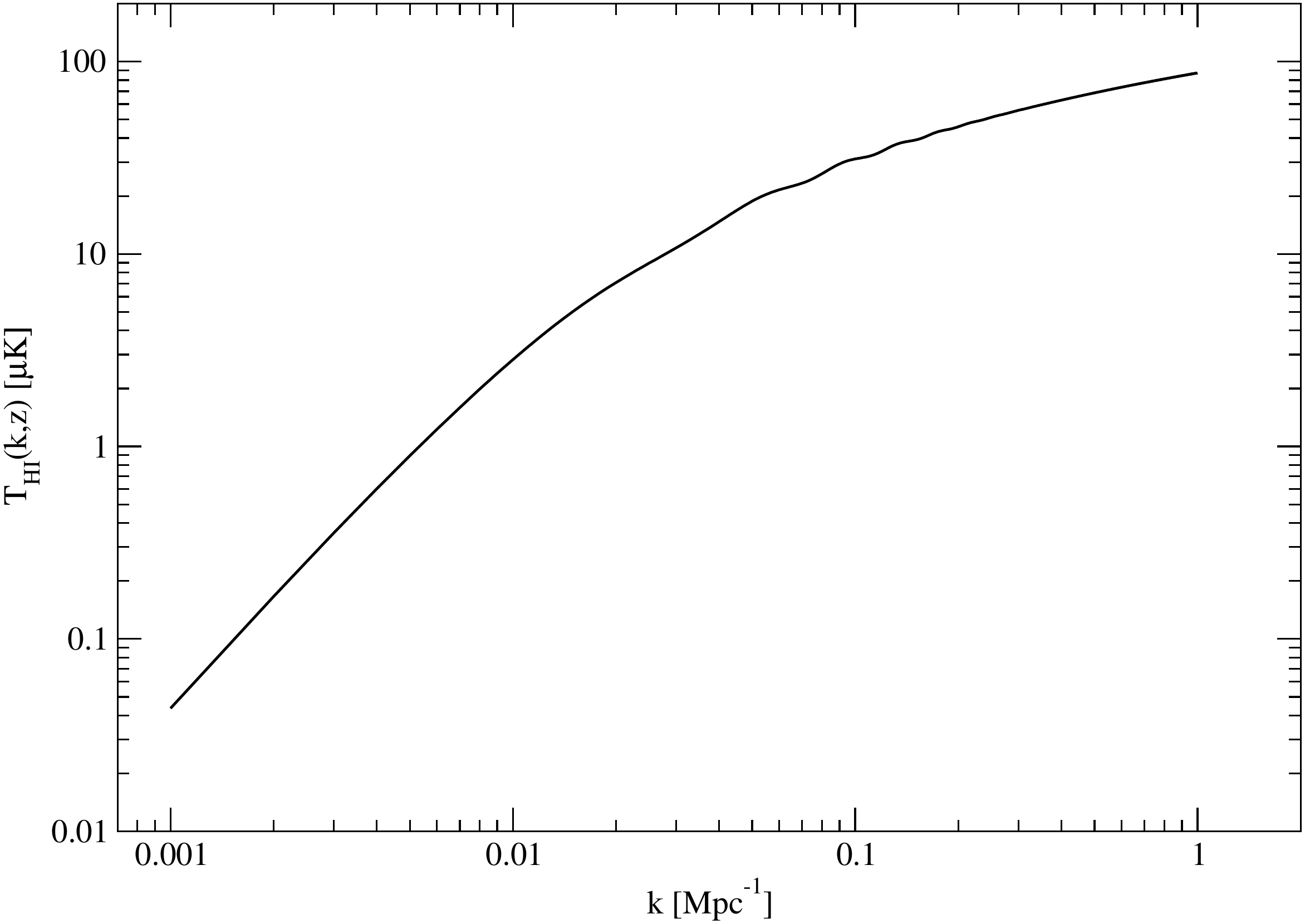}
\caption{The 3D HI power spectrum at $z=0.28$. BAOs are just about visible in the range $0.01<k/ \rm{Mpc}^{-1}<0.2$, but they are less than $10\%$ of the total signal. For $k<0.01\,{\rm Mpc}^{-1}$ the temperature spectrum is $\propto k^2$ corresponding to $P(k)\propto k$, that is, scale invariant.}
\label{fig:PHI3Dzeq05}
\end{figure}

The 3D HI power spectrum can be written as
 \be
\label{eq:hatP}
[\Delta T_{\rm HI}(k,z)]^2=\bar{T}(z)^2 [b(k,z)]^2 {k^3P_{\rm{cdm}}(k,z)\over 2\pi^2}, \ee where $b(k,z)$ is
the bias and $P_{\rm{cdm}}$ is the
underlying dark matter power spectrum.  We will assume, for the purposes of this discussion, that  $b=1$ independent of redshift, $z$, and scale $k$.  \citet{Martin:2012} find that this is a good assumption for scales $>10h^{-1}\,{\rm Mpc}$ implying that, within the constraints of their survey, that HI galaxies trace the large-scale structure of the Universe. In Fig.~\ref{fig:PHI3Dzeq05} we plot the predicted 3D linear power spectrum of 21 cm fluctuations for 
$z =0.28$ which corresponds to the case (A) discussed in section~\ref{sec:optimise} - the spectrum at other values of $z$ is virtually indistinguishable on this logarithmic scale except for the overall growth factor. At low $k$, large angular scales, the signal has a spectrum $T_{\rm HI}\propto k^2$, which turns over around $k\sim 0.1\,{\rm Mpc}^{-1}$. BAOs are less than $10\%$ of the signal in the range $0.01<k/ \rm{Mpc}^{-1}<0.2$ and the power spectrum is significantly affected by the non-linear regime for $k>k_{\rm nl}$ which is $\approx 0.2\,{\rm Mpc}^{-1}$ at $z=0$. The wavenumber $k$ can be converted to angular multipole using $\ell\approx kr(z)$ where $r(z)$ is the coordinate distance and this can in turn be converted to an approximate angle scale using $\theta\approx 180\,{\rm deg}/\ell$. The non-linear scale corresponds to $\ell_{\rm nl}\approx 220$ and $\theta_{\rm nl}\approx 0.8\,{\rm deg}$ at low redshift $(z\approx 0.3)$. Observations which probe significantly higher resolution than this will require the modelling of non-linear effects on the power spectrum. Note that this is somewhat different at higher redshifts.

\subsection{2D Angular power spectrum}
\label{sec:appCl}

One can construct a 2D angular power spectrum for the HI intensity over
some frequency range which can then be treated in using techniques designed to study the CMB
anisotropies. We start from the 3D quantity
$\delta\tilde{T}(r(z)\bn,z)$ which can be computed directly from
(\ref{eq:Tobs1}) by replacing $\rho_{\rm HI}$ with $\delta\rho_{\rm
HI}$. This makes the assumption that $d\ell/d v$ and $1+z$ take the
values in an unperturbed universe, that is, we ignore the effects of
peculiar velocities, which will lead to redshift space distortions
that could be important on small scales, and the Sachs-Wolfe effects
which could be important on large scales~\citep{Challinor:2011,Bonvin:2011}. Its projection on the sky
$\delta T(\bn)$ is defined using a normalized projection kernel
$W(z)$, that is
\begin{eqnarray}
\delta T(\bn)&=&\int dz W(z)\delta\tilde{T}(r(z)\bn,z)\nonumber\\
&=&\int dz W(z)\bar{T}(z)\delta_{HI}(r(z)\bn,z).
\end{eqnarray}
We will use a top-hat projection kernal
$$
W(z) = \left\{ \begin{array}{rl}
\frac{1}{z_{\rm{max}}-z_{\rm{min}}} &\mbox{ if $z_{\rm{min}}\leq z \leq z_{\rm{max}}$} \\
0 &\mbox{ otherwise,}
\end{array} \right.
$$
but the concept is more general and any weighting function could be used. The maximum and minimum redshifts of the bin are related to the minimum and maximum frequencies of the bin.

Fourier transforming and then expanding the Fourier
modes in spherical harmonics, we have
\bea
\delta\tilde{T}(r(z)\bn,z)&=&\bar{T}(z)\int \frac{d^3k}{(2\pi)^3}\hat{\delta}_{HI}(\bk,z)e^{ir(z)\bn \bk} \nonumber \\
&=& 4\pi \bar{T}(z)\sum_{l,m} i^{\ell}\int \frac{d^3k}{(2\pi)^3}\hat{\delta}_{HI}(\bk,z)
 j_{\ell}(kr(z))\nonumber \\ &\times&Y^{*}_{\ell m}(\hat{\bk})Y_{\ell m}(\bn)\,,
\eea
where  $j_{\ell}$ is the spherical Bessel function. We then have
\bea
\delta T(\bn)&=&4\pi \sum_{l,m} i^{\ell} \int dz W(z) \bar{T}(z) \nonumber\\
&\times& \int \frac{d^3k}{(2\pi)^3}\hat{\delta}_{HI}(\bk,z)  j_{\ell}(kr(z))Y^{*}_{\ell m}(\hat{\bk})Y_{\ell m}(\bn)\,,
\eea
which gives
\be
a_{\ell m}=4\pi i^{\ell}\int dz W(z)\bar{T}(z)\int \frac{d^3k}{(2\pi)^3}\hat{\delta}_{HI}(\bk,z)  j_{\ell}(kr(z))Y^{*}_{\ell m}(\hat{\bk}).
\ee

The angular power spectrum $C_{\ell}$ is found by taking the ensemble average
\be
C_{\ell} \equiv \langle a_{\ell m} a^{*}_{\ell m}\rangle.
\ee
We need to use the orthonormality condition for the spherical harmonics, statistical isotropy and the definition of the power spectrum
\be
\langle \hat{\delta}_{HI}(\bk,z) \hat{\delta}_{HI}(\bk',z') \rangle = (2\pi)^3 \delta (\bk-\bk')b^2P_{\rm cdm}(k)D(z)D(z').
\ee
Here, $D(z)$ is the growth factor  for dark matter perturbations defined such that $D(0)=1$. Then we find that
\begin{eqnarray}
C_{\ell} = \frac{2b^2}{\pi}\int dz W(z)\bar{T}(z)D(z) \int dz' W(z')\bar{T}(z')D(z') \nonumber\\\times\int k^2 dk P_{\rm cdm}(k)j_{\ell}(kr(z))j_{\ell}(kr(z')).
\end{eqnarray}

For large $\ell$, we can use the Limber approximation~\citep{Limber:1953,Afshordi:2008} to perform the $k$ integral, that is,
\begin{eqnarray}
\left(\frac{2}{\pi}\right)\int k^2dkP_{\rm cdm}(k)j_{\ell}(kr)j_{\ell}(kr') \\
=P_{\rm cdm}\left(\frac{\ell + 1/2}{r}\right)\frac{\delta(r-r')}{r^2}\,.
\end{eqnarray}
 Writing $cdz'=H_0E(z^{\prime})dr'$, we can use the $\delta$-function to perform the $dr'$ integral and we deduce
\be
\label{cl}
C_{\ell}=\frac{H_0b^2}{c} \int dz E(z) \left[{W(z) \bar{T}(z)D(z)\over r(z)}\right]^2 P_{\rm cdm}\left(\frac{\ell+\frac{1}{2}}{r}\right).
\ee

We can proceed to calculate the $21$cm angular power spectrum using
(\ref{cl}).  The matter power spectrum $P_{\rm cdm}(k)$ today
($z=0$) can be computed from \be P_{\rm{cdm}}(k)=Ak^{n_{\rm S}}T^2(k).  \ee
For the transfer function $T(k)$ we use the fitting formulae given by
\citet{Eisenstein:1997ik}, assuming
$\Omega_{\rm{m}}=0.27$,
$\Omega_{\Lambda}=1-\Omega_{\rm{m}}$, $h=0.71$ and baryon fraction $f_{\rm{baryon}}=0.17$. We take
the primordial spectral index to be $n_{\rm S}=0.96$ and use the approximate
formula by \citet{Bunn:1996da} for the power spectrum
amplitude. For this choice of parameters, the sound horizon $s \approx
152 \, \rm{Mpc}$ and the BAO scale $k_A = 2\pi/s \approx 0.041 \,
\rm{Mpc}^{-1}$. The growth function $D(z)$ is calculated using the
fitting formula from  \cite{Carroll:1991mt}.

\begin{figure}
\centering
\includegraphics[scale=0.3]{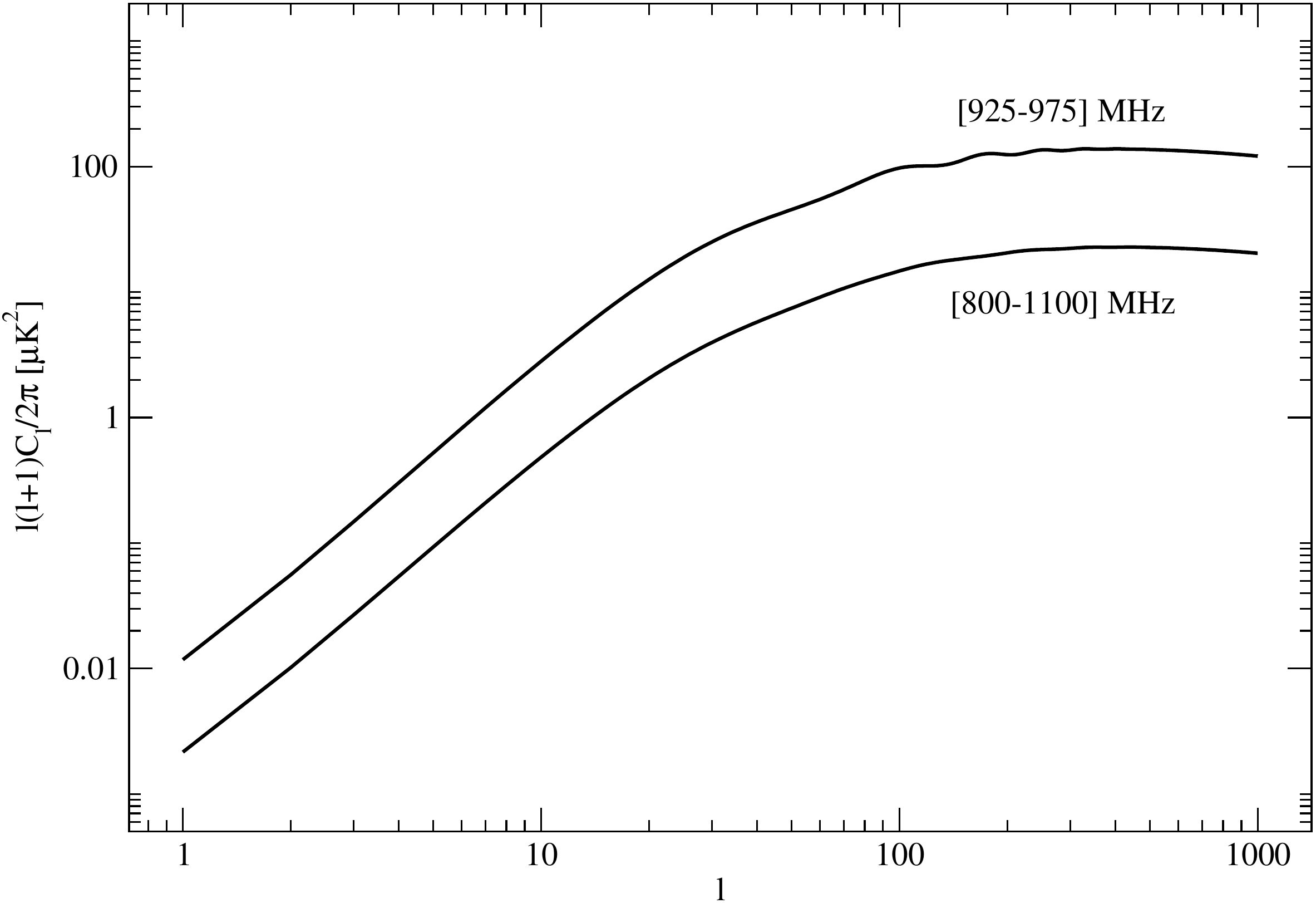}
\caption{The HI angular power spectrum for the same central frequency, $950 \, \rm{MHz}$, but for two different ranges $925$--$975\,{\rm MHz}$ and $800$--$1100\,{\rm MHz}$. The signal amplitude increases with decreasing the frequency range and the BAOs become more prominent. As the frequency range increases the size of the region probed becomes larger implying that the fluctuations will decrease.}
\label{Fig:HIspec}
\end{figure}

\begin{figure}
\centering
\includegraphics[scale=0.3]{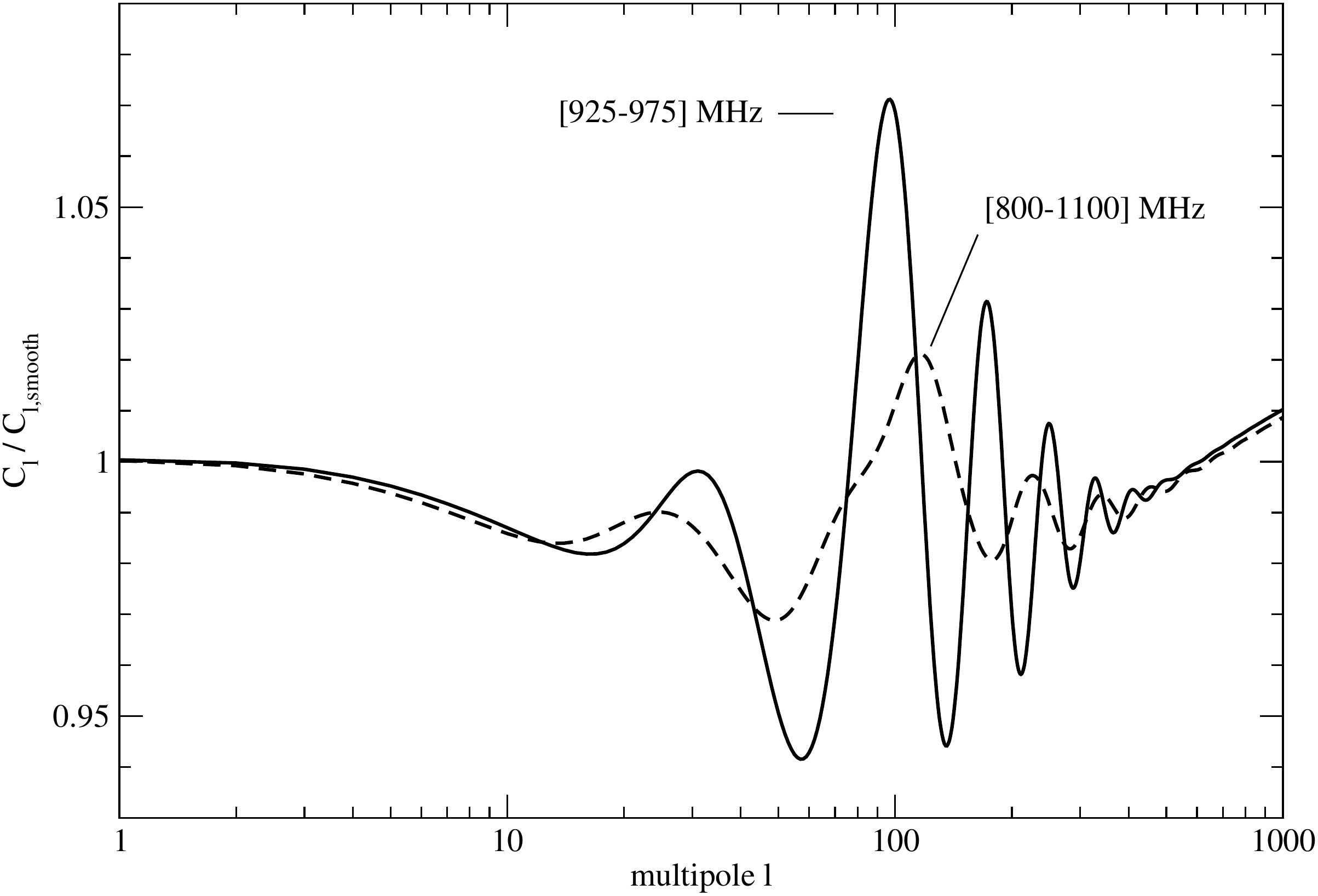}
\includegraphics[scale=0.3]{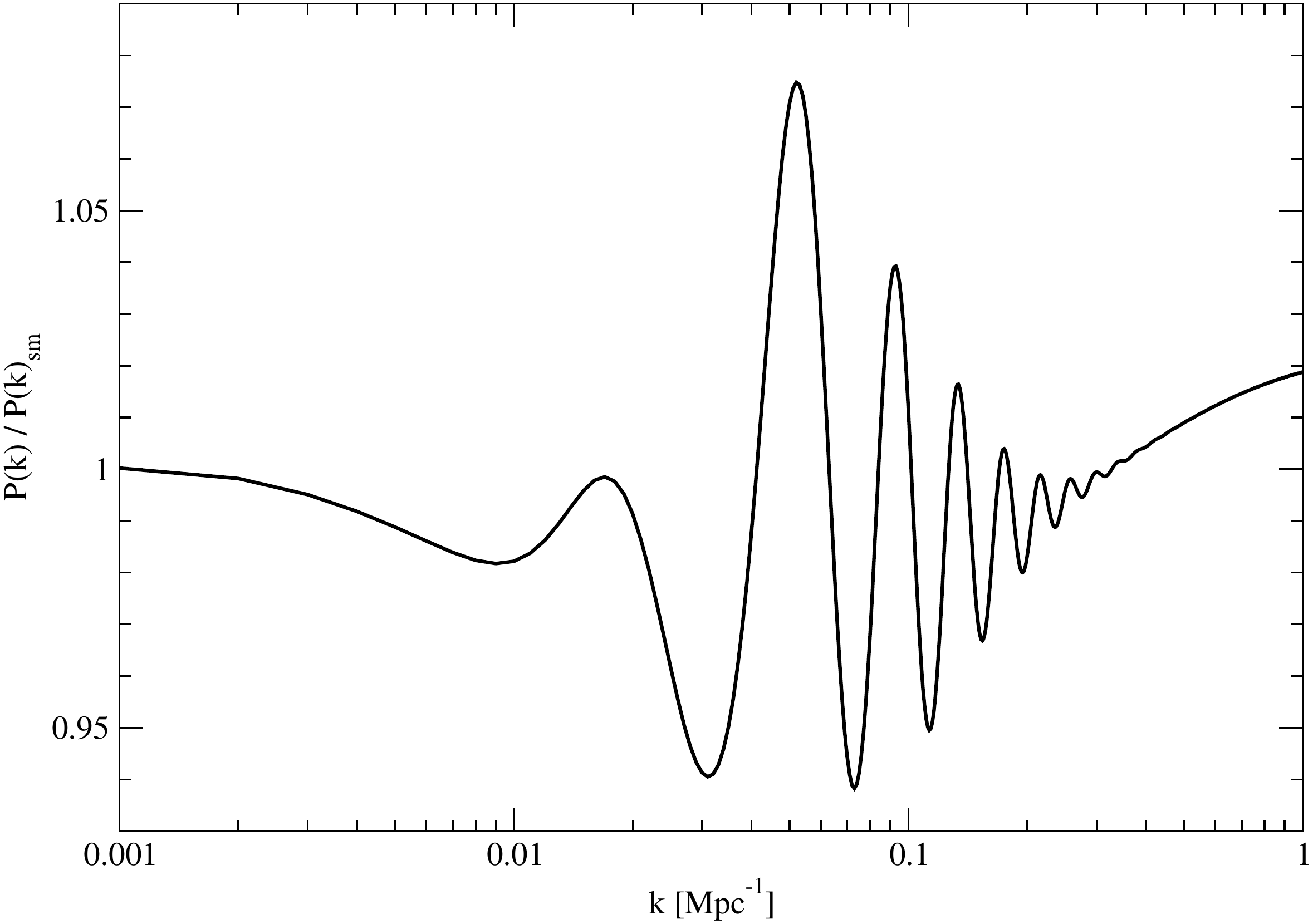}
\caption{In the top plot we present the ratio $C_{\ell}$/$C_{\ell,\rm{smooth}}$ for the same central frequency ($950 \, \rm{MHz}$) but for different frequency ranges. In the bottom we present the ratio $P(k)/P(k)_{\rm{smooth}}$. The $925-975\,{\rm MHz}$ curve in the upper plot is very similar to that in the lower one implying that a frequency range of $50\,{\rm MHz}$ will be optimal to detect the BAO signal at this redshift.}
\label{PkovPksm}
\label{Fig:ratios}
\end{figure}

In Fig.~\ref{Fig:HIspec} we plot the HI angular power spectra for the
same central frequency $950 \, \rm{MHz}$ (corresponding to a central
redshift $z_c \approx 0.5$) but for two different bin widths $\Delta f
=300 \, \rm{MHz}$ and $\Delta f =50 \, \rm{MHz}$. We see that as a
result of decreasing the bin size, the signal amplitude increases and
the BAO wiggles become more prominent. It is clear that there exists
an optimum frequency bin size for which a detection will be possible: the
signal-to-noise level would increase $\propto \sqrt{\Delta f}$ if the
signal level was independent of frequency, but the level of
fluctuations in the HI is reduced as the box enclosed by the angular
beam-size and frequency range are increased. In Fig.~\ref{Fig:ratios}
we plot the ratio $C_{\ell}$/$C_{\ell,\rm{smooth}}$, where
$C_{\ell,\rm{smooth}}$ is the no-wiggles angular power spectrum
\citep{Eisenstein:1997ik} for the same central frequency $950 \,
\rm{MHz}$ but for the two frequency ranges used in
Fig.~\ref{Fig:HIspec}. In addition we have plotted the ratio
$P(k)/P_{\rm{smooth}(k)}$, where $P(k)=P_{\rm{cdm}}(k)$. Note that the
ratio $C_{\ell}$/$C_{\ell,\rm{smooth}}$ cannot be bigger than
$P(k)/P_{\rm{smooth}(k)}$. However, the two are in close agreement for
the $\Delta f =50 \, \rm{MHz}$ bin which indicates that this is the
optimal frequency bin size for tracing the BAOs.

\subsection{Estimated signal from ${\rm S}^3$ simulations}

\citet{Obreschkow:2009a, Obreschkow:2009b} present semi-analytic simulations of HI emission based on catalogues of galaxies whose properties are evolved from the Millennium or Milli-Millennium dark matter simulations \citep{Springel:2005} and we will use this to compute simulated spectra as would be observed by a single dish. We adopt the smaller of the two resultant mock sky fields (corresponding to the Milli-Millennium simulation, comoving diameter $s_{\mathrm{box}}=62.5\, h^{-1}\,{\rm Mpc}$) for the characterisation of the expected HI signal which, as we will discuss later, restricts the highest redshift to which the spectrum will be complete. We extract galaxy properties from the simulated catalogue - position coordinates, redshift, HI mass, integrated HI flux and HI-line width - from which we calculate the HI brightness temperature contributed by galaxies in the frequency range $\Delta f$, as it would be observed by a telescope  beam which is a top hat of diameter $\theta_\mathrm{FWHM}$ independent of frequency.  Using a diffraction limited beam for which $\theta_{\rm FWHM}\propto 1+z$ would result in a higher signal at lower frequencies.

\begin{figure}
     \centering
     \subfigure[$\, \Delta f = 1 \, \rm{MHz}, \theta_{\rm{FWHM}} = 60 \, \rm{arcmin}$]{
              \includegraphics[width=.35\textwidth]{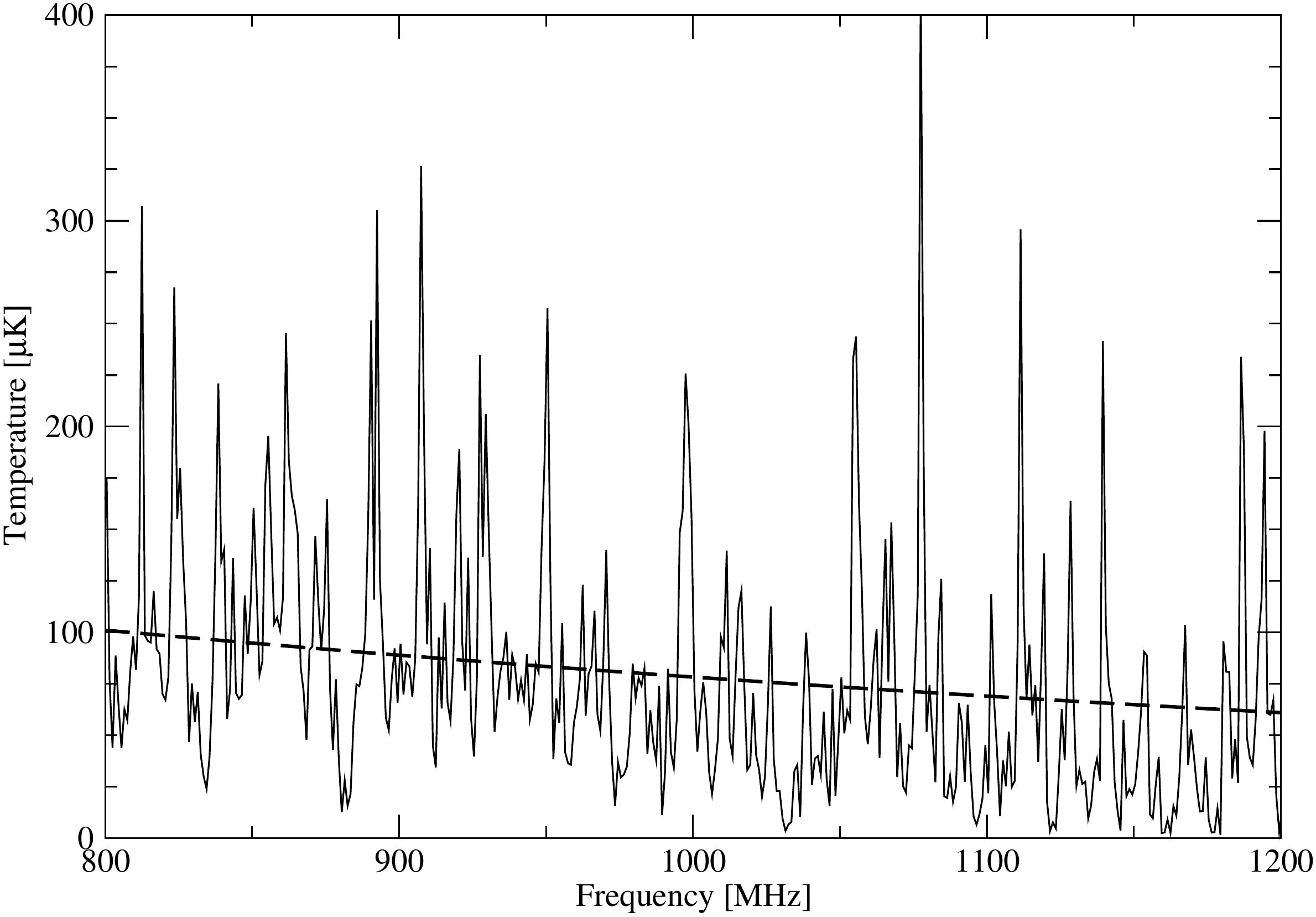}}
     \subfigure[$\, \Delta f = 10 \, \rm{MHz}, \theta_{\rm{FWHM}} = 60 \, \rm{arcmin}$]{
          \includegraphics[width=.35\textwidth]{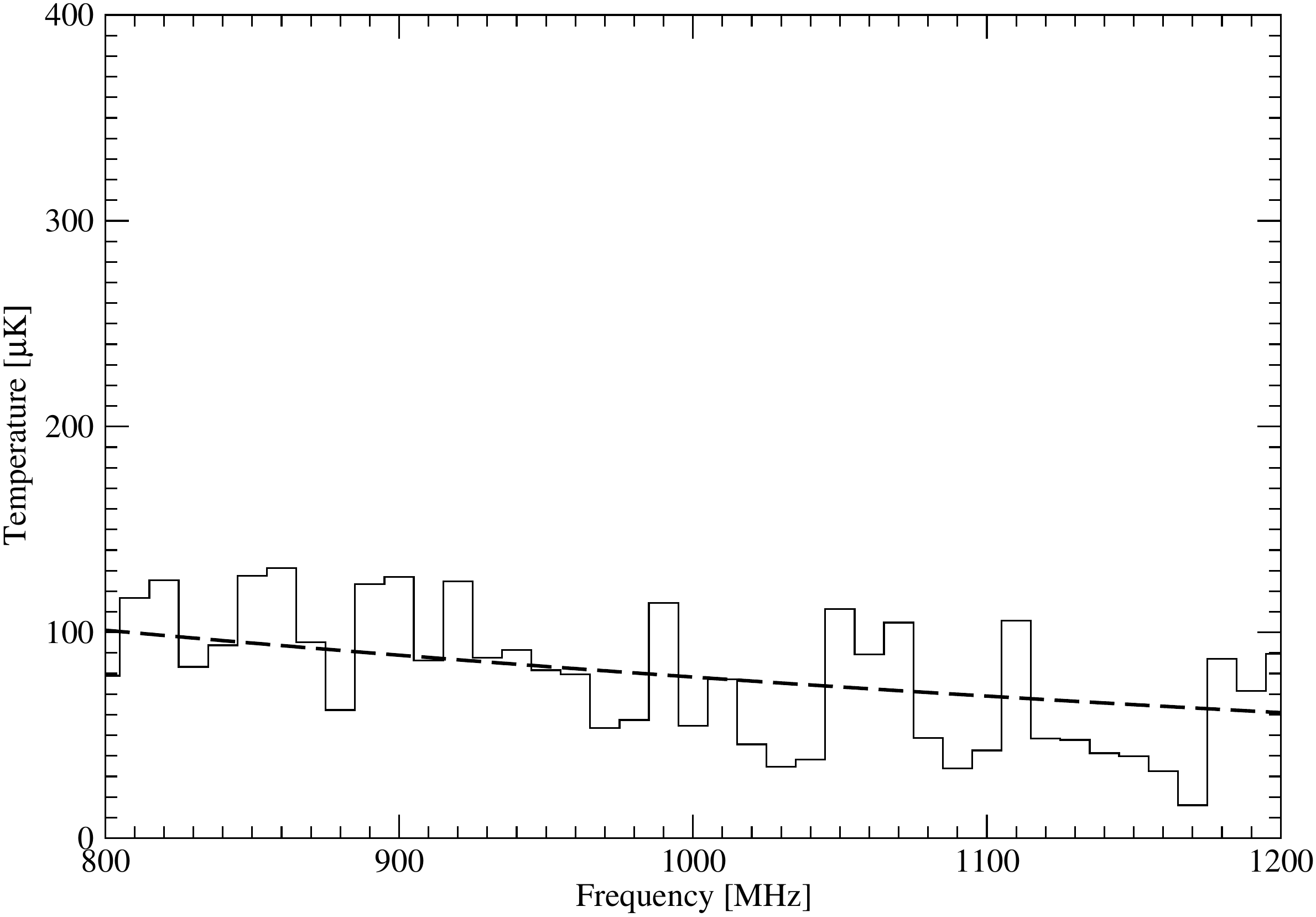}}\\
     \subfigure[$\, \Delta f = 1 \, \rm{MHz}, \theta_{\rm{FWHM}} = 20 \, \rm{arcmin}$]{
                \includegraphics[width=.35\textwidth]{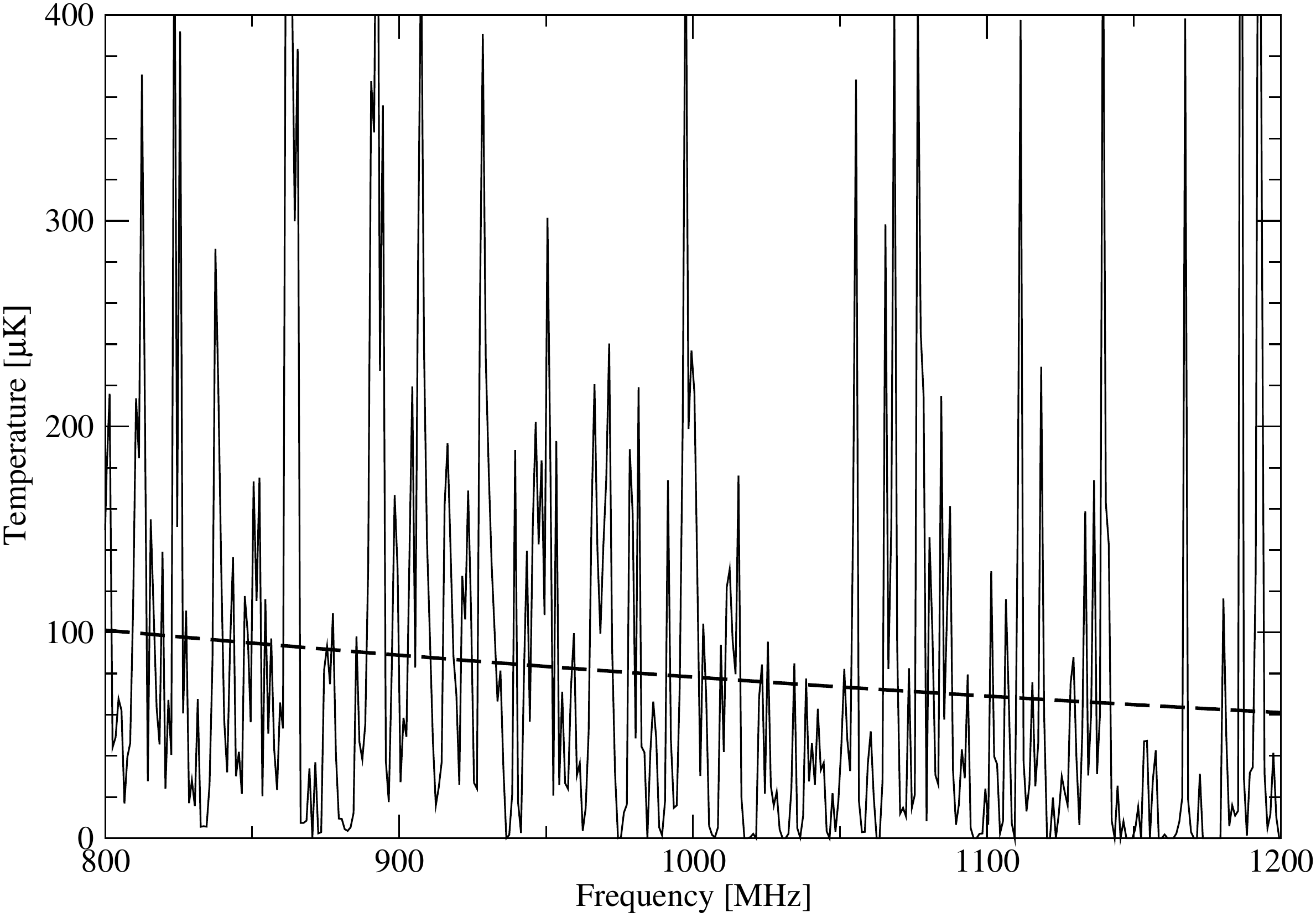}}
     \subfigure[$\, \Delta f = 10 \, \rm{MHz}, \theta_{\rm{FWHM}} = 20 \, \rm{arcmin}$]{
          \includegraphics[width=.35\textwidth]{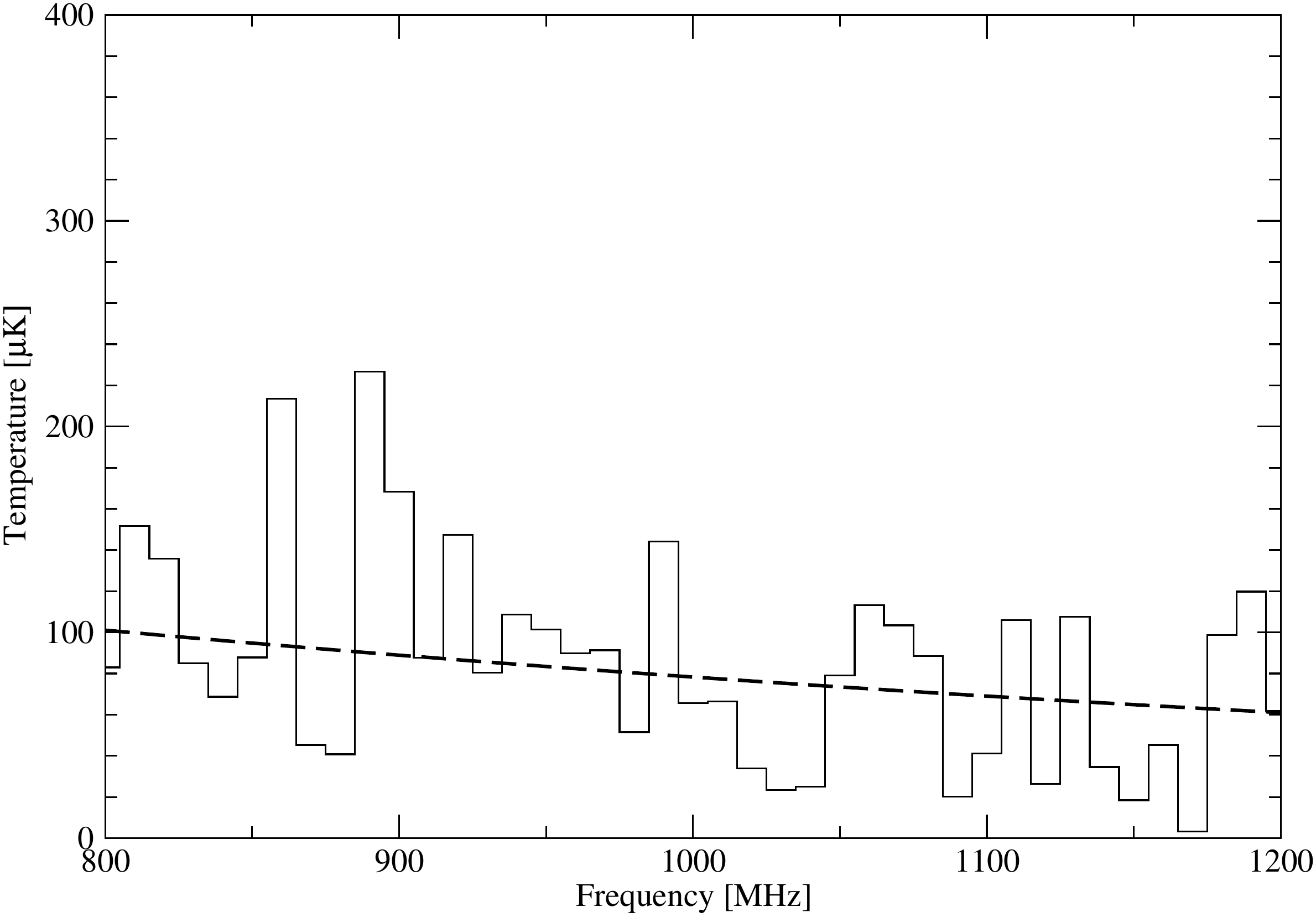}}
     \caption{The simulated HI signal for different telescope beam resolution and frequency
     bin size combinations, $\theta_{\rm FWHM}$ and $\Delta f$ respectively. A smaller frequency bin or telescope beam size results in larger
     fluctuations, as expected, since going to smaller values of $\theta_{\rm FWHM}$ and $\Delta f$ means that one is probing a smaller volume of the Universe and fluctuations in the HI density are expected to be larger on smaller scales, as they are in the cold dark matter. In each case the dashed line is the average temperature computed using (\ref{tave}). The objective of 21cm intensity mapping experiments is to measure the difference between the signal and this average.}
     \label{tempfreqdfdtheta}
\end{figure}

In Fig.~\ref{tempfreqdfdtheta} we present the expected HI temperature fluctuations between frequencies
of $800$ and $1200\,{\rm MHz}$, which is part of the frequency range we will be interested in, for a range of beam sizes, $\theta_{\rm FWHM}$ and frequency bandwidths, $\Delta f$. 
These plots provide considerable insight into the nature of the signal that we will be trying to detect. It appears that the average signal from equation (\ref{tave}), represented by the dashed line, is compatible with the spectra from the Millennium Simulation. Larger values of $\theta_{\rm FWHM}$ and $\Delta f$ mean that the volume of the universe enclosed within a single cell is larger and hence the spectra have only very small deviations from the average, whereas for smaller values there are significant deviations as large as $400\,\mu{\rm K}$ illustrating that the signal on these scales can be much higher than the average.

There are limitations to this approach: our spectra are not accurate out to arbitrarily high redshift where the cone diameter exceeds $s_\mathrm{box}$ and hence simulated catalogue is missing galaxies, that is, it is not complete. The maximum redshift for which the simulated catalogue is complete is determined by the beam size, with smaller beam sizes being accurate to higher redshifts.  For a beam size of $1^{\circ}$, the spectrum will be accurate up to around $600\,{\rm MHz}$.

\begin{figure}
\centering
\includegraphics[scale=0.3]{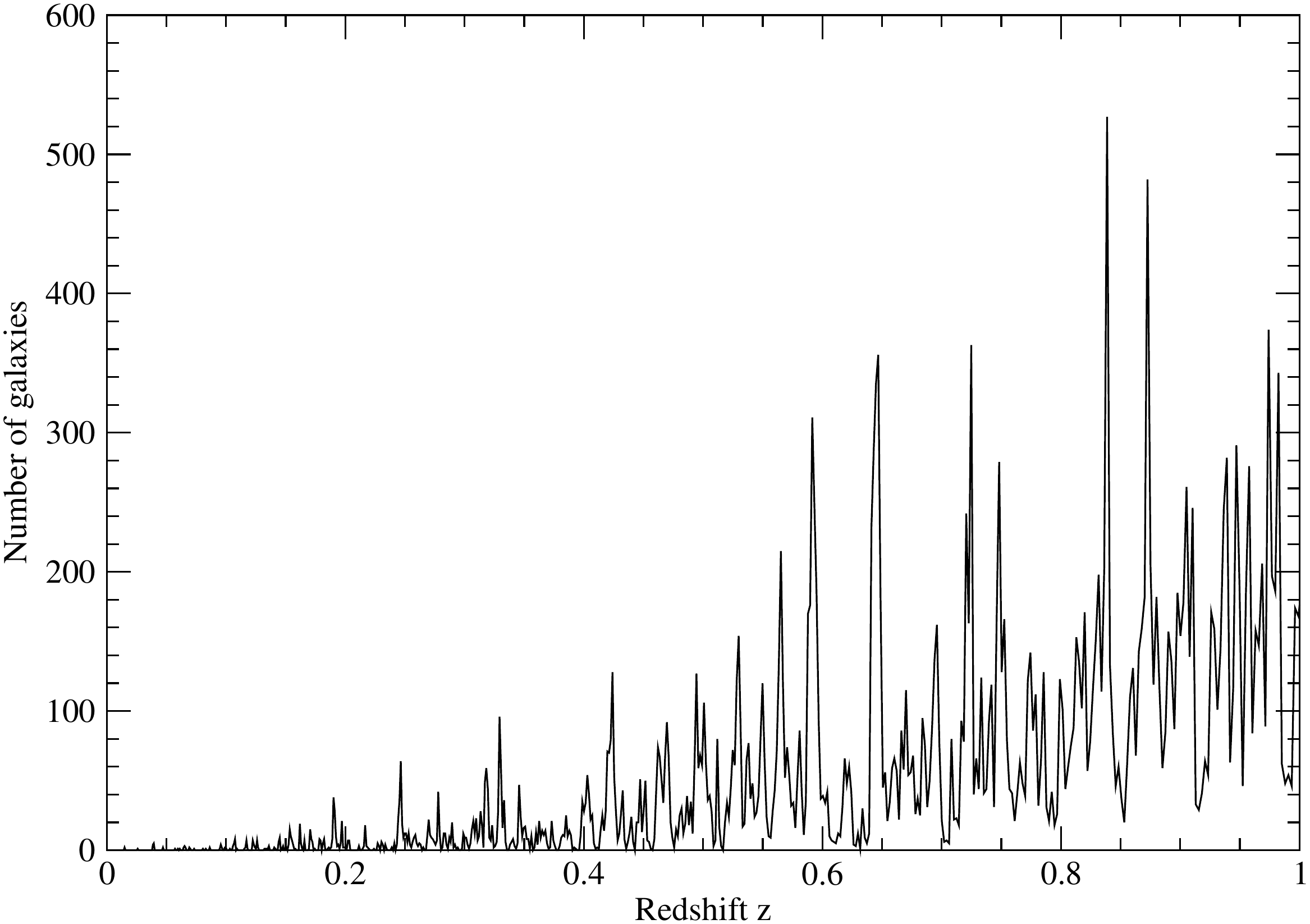}
\caption{The number of galaxies contributing to the HI signal in each bin as a function of redshift. We have used $\theta_{\rm FWHM}=20\,{\rm arcmin}$ and $\Delta f=1\,{\rm MHz}$ for this particular example. Note that the signal presented in Fig.~\ref{tempfreqdfdtheta}© varies only very weakly with redshift yet the number of galaxies contributing to each bin increases dramatically. At low redshifts, we see that the signal is dominated by individual galaxies, at intermediate redshifts it is dominated by galaxy clusters and groups as suggested in \citet{Battye:2004} and at high redshifts the signal comprises an unresolved background signal, that is, there are a very large number of galaxies contributing to each bin in redshift.}
\label{Ngal}
\end{figure}

At the other end of the frequency range, which is not shown in Fig.\ref{tempfreqdfdtheta}, there are large fluctuations around the
mean that  can be attributed to the effects of shot noise. In
Fig.~\ref{Ngal}, we plot the number of galaxies contributing to each bin of the spectrum defined by $\theta_{\rm FWHM}=20\,{\rm arc min}$ and  $\Delta f=1\,{\rm MHz}$, that is, Fig.~\ref{tempfreqdfdtheta}(c),  as a function of redshift. At low redshifts (high
frequencies) the number of   galaxies contributing to each bin is small, and the signal cannot be
thought of as an unresolved background. If $N$ is the number of contributing galaxies, the
deviation due to shot noise is $\sigma^2 \propto 1/N$ and
consequently, we see larger fluctuations. This is exactly the regime probed by surveys such as HIPASS and ALFALFA which detect individual galaxies for $z<0.05$ with slightly better - but not substantially better - resolution than used in this plot. As one moves to higher redshifts a low resolution telescope will preferentially detect objects which have an angular size which is similar to that of the beam. At intermediate redshifts this will be the clusters discussed in \citet{Battye:2004} and these can be clearly seen as the well-defined spikes in the spectrum around $z=0.3-0.4$. At higher redshifts even clusters are beam-diluted; beyond $z=0.5$ there are always a few galaxies contributing to each bin - essentially there is an unresolved background.

\begin{figure}
\centering
\includegraphics[width=0.45\textwidth]{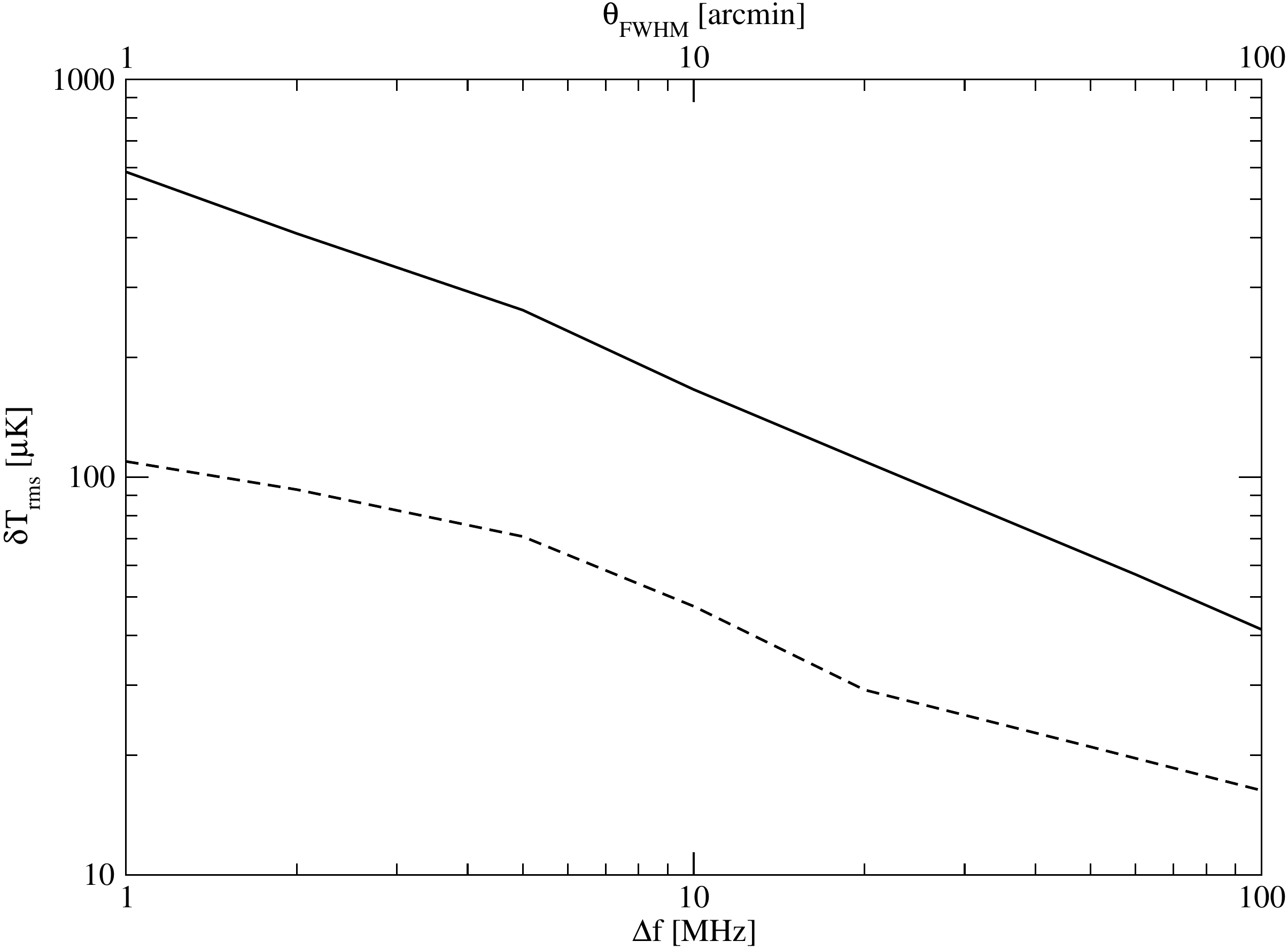}
\caption{The r.m.s. temperature fluctuations in the HI signal as a function of $\theta_{\rm{FWHM}}$  (solid line) and $\Delta f$ (dashed line). When we vary $\theta_{\rm FWHM}$ we fix $\Delta f=1\,{\rm MHz}$ and in the case where $\Delta f$ is varied we fix $\theta_{\rm FWHM}=20\,{\rm arcmin}$.}
\label{rmsdfdtheta}
\end{figure}

We can quantify the HI temperature fluctuations as a function of $\theta_{\rm{FWHM}}$ and $\Delta f$ by
calculating their root mean squared (r.m.s.) deviation about the mean over the frequency range. As shown in Fig.~\ref{rmsdfdtheta}, the r.m.s. decreases with increasing $\theta_{\rm{FWHM}}$ and $\Delta f$. This is exactly what is expected since as $\theta_{\rm{FWHM}}$ and $\Delta f$ increase the region of the universe enclosed is larger and the observed temperature will become closer to the average. We note that typically the fluctuations are significantly bigger than one for a substantial part of the parameter space.
 
\begin{figure}
\centering
\includegraphics[scale=0.4]{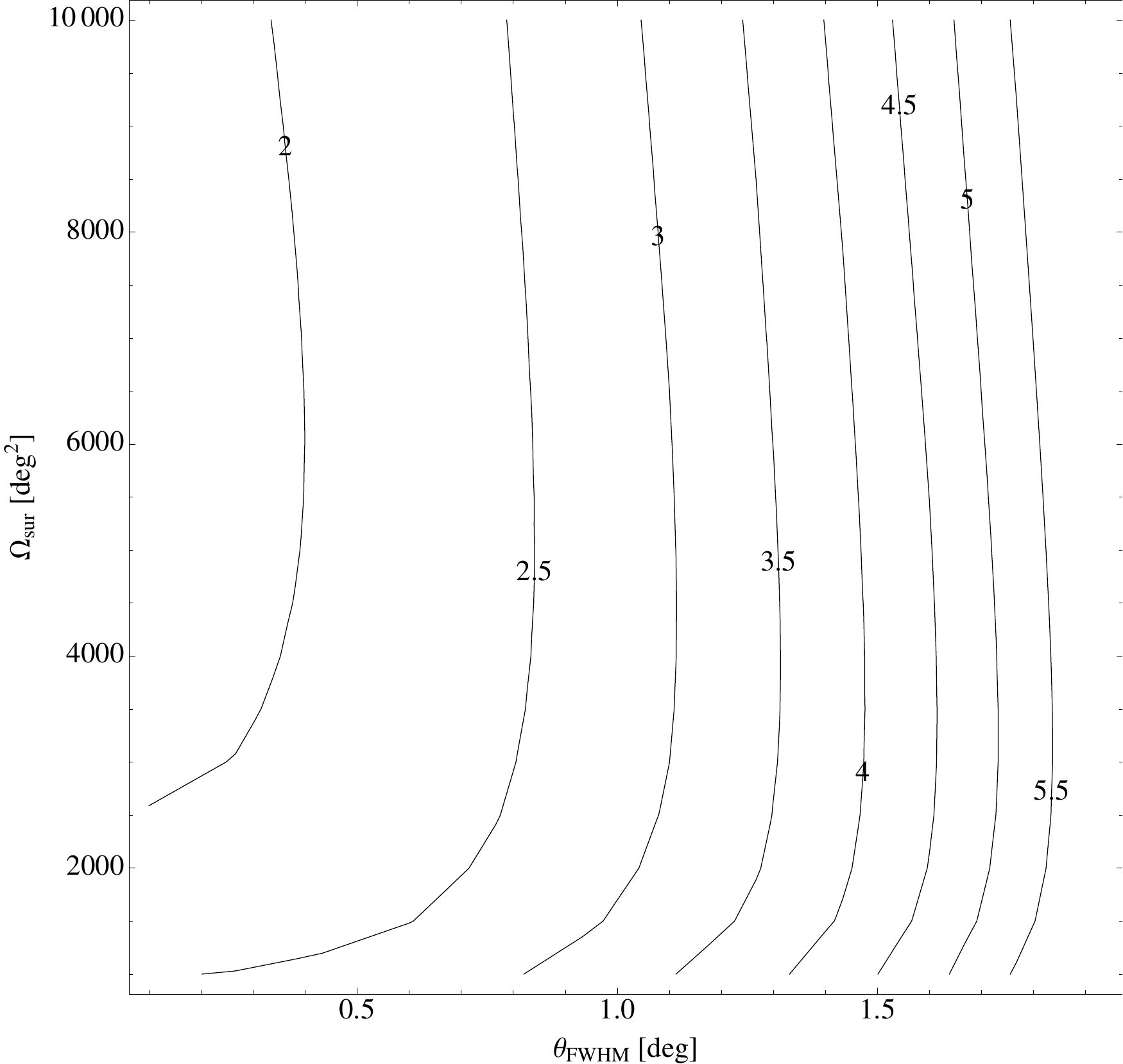}
\includegraphics[scale=0.4]{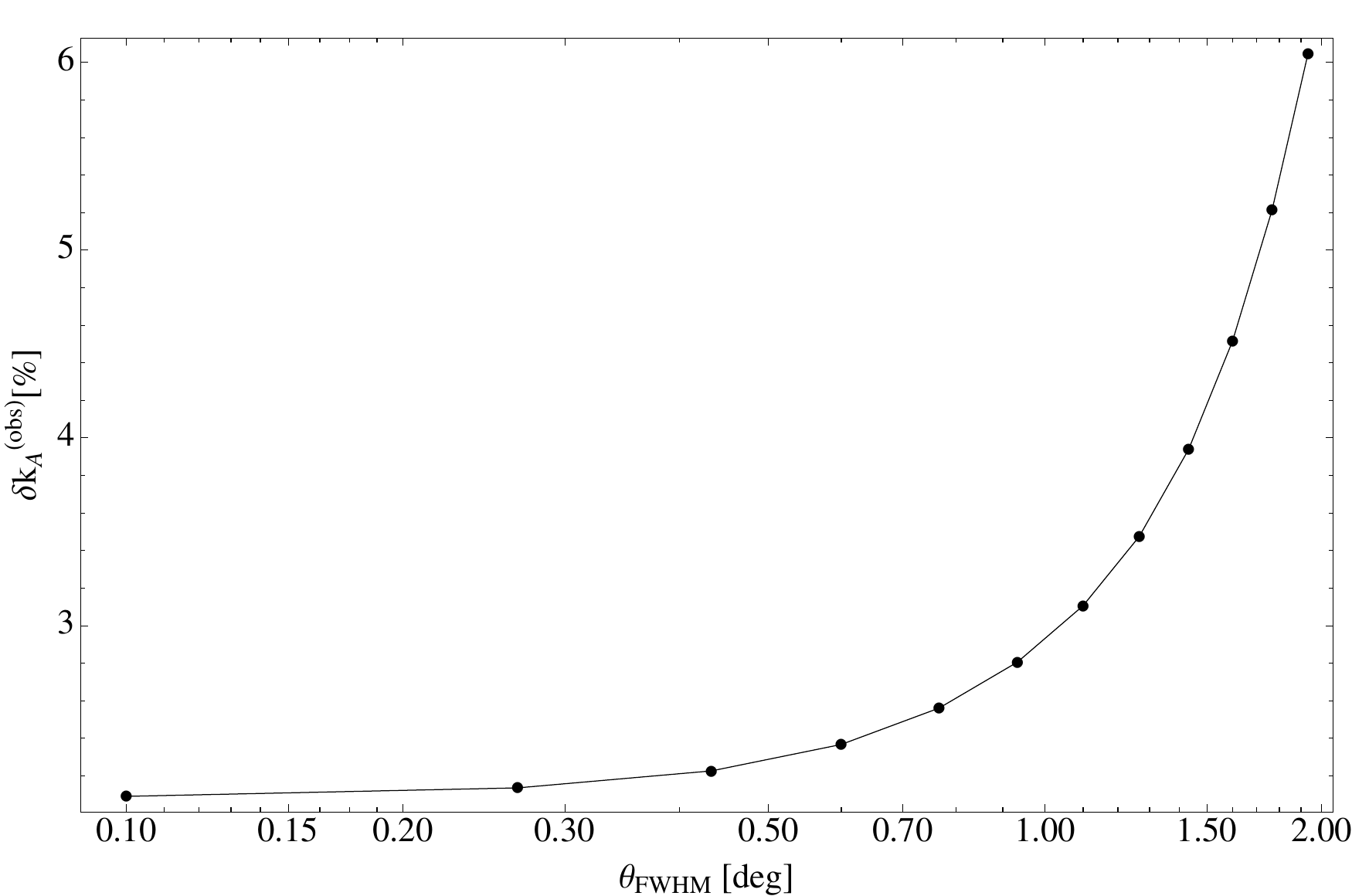}
\includegraphics[scale=0.4]{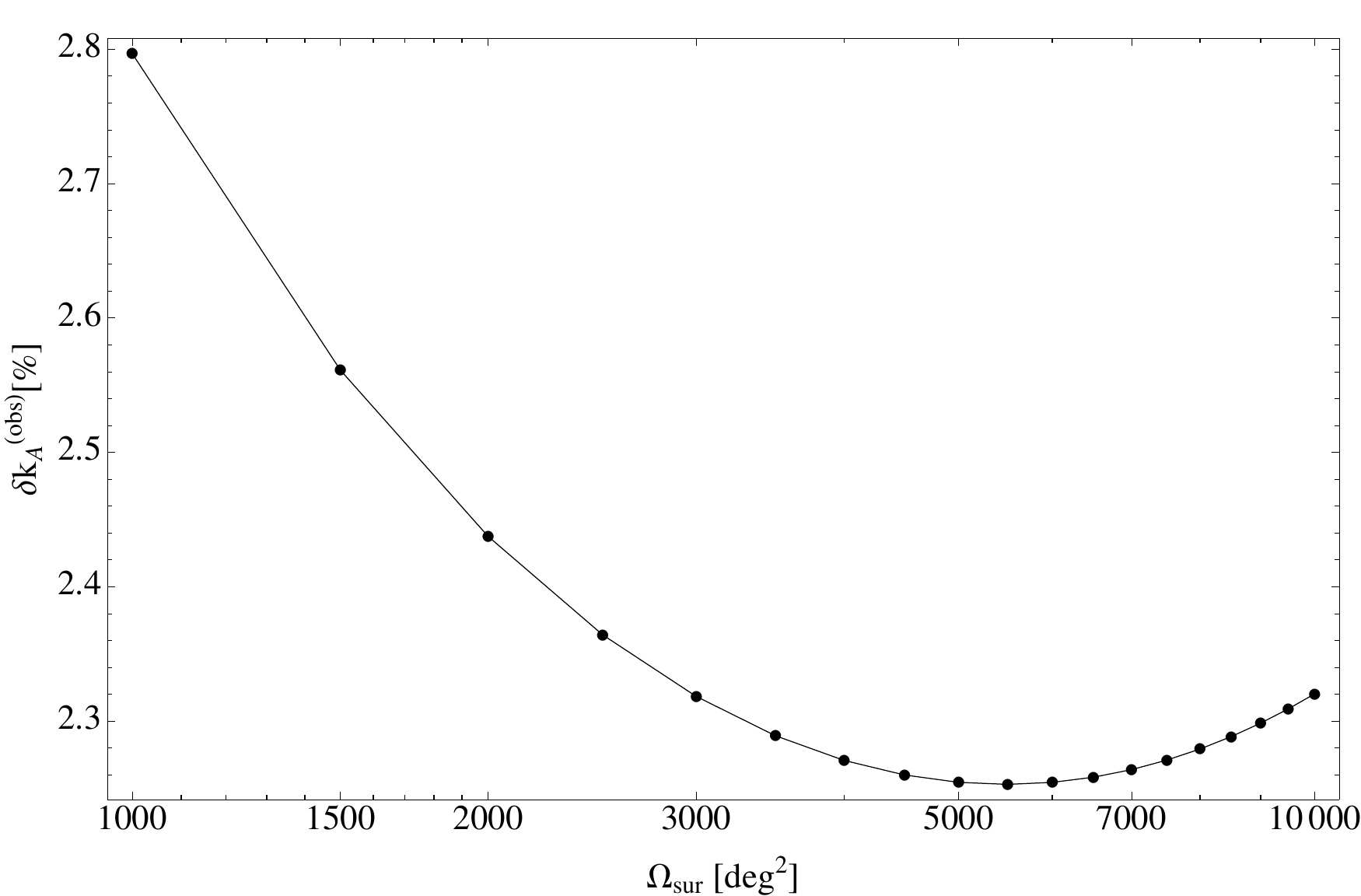}
\caption{Optimization of the measurement of $k_{\rm A}$ for case (A) - observations over the frequency range $960-1260\,{\rm MHz}$. We have computed the projected uncertainty on the measurement of $k_{\rm A}$ as a function of $\Delta\Omega_{\rm sur}$ and the beam size $\theta_{\rm FWHM}$.  In the top panel the contours are lines of constant $\delta k_{\rm A}$ with the numbers on the contours signifying the percentage fractional error: $100\delta_{\rm A}/k_{\rm A}$.  In the middle panel we present a 1D slice as function of  $\theta_{\rm FWHM}$ with $\Delta\Omega_{\rm sur}=2000\,{\rm deg}^2$ and in the bottom panel a 1D slice as a function of $\Omega_{\rm sur}$ with $\theta_{\rm FWHM}=40\,{\rm arcmin}$.}
\label{fig:ratio3D_A}
\end{figure}

\begin{figure}
\centering
\includegraphics[scale=0.4]{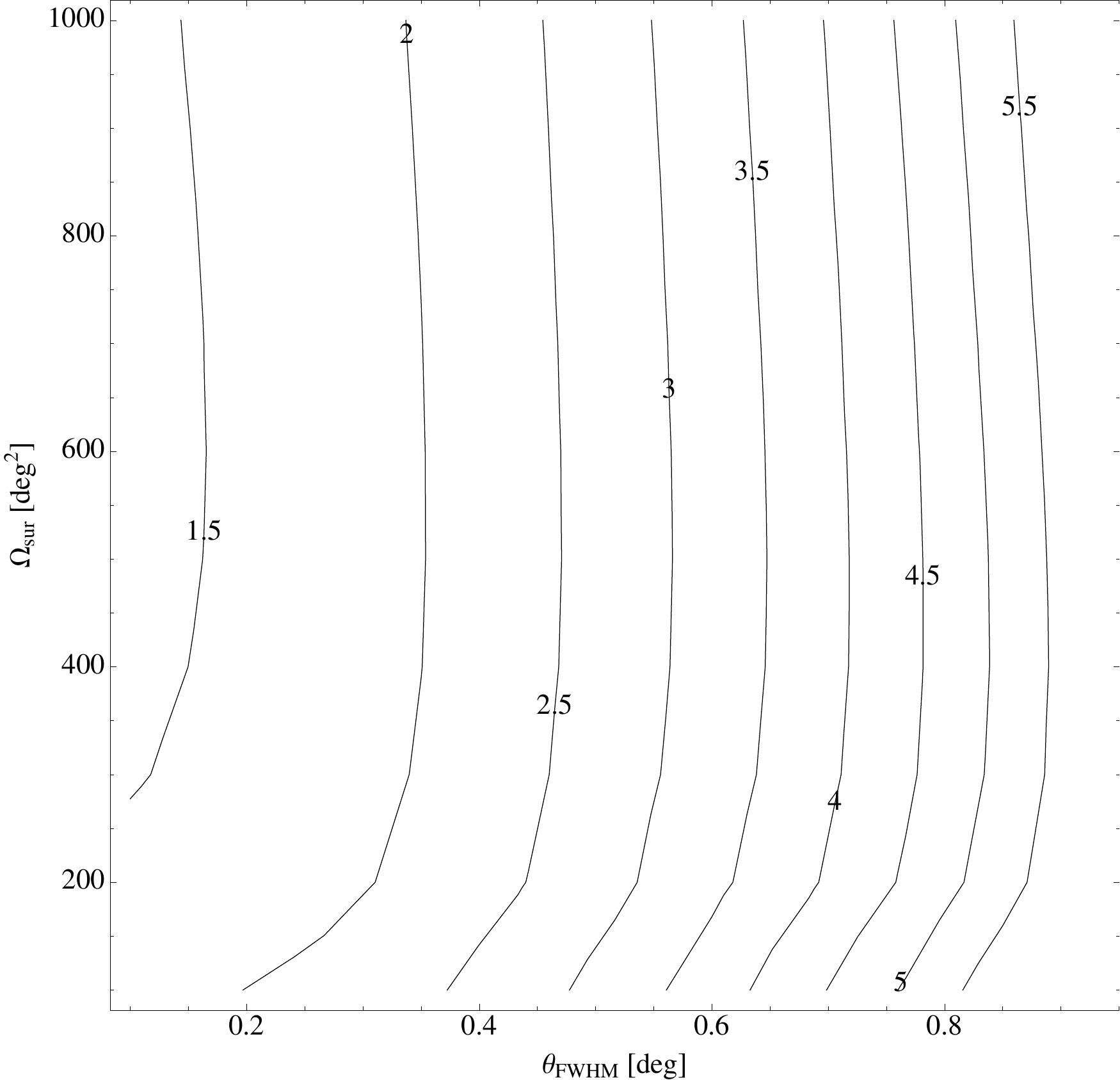}
\includegraphics[scale=0.4]{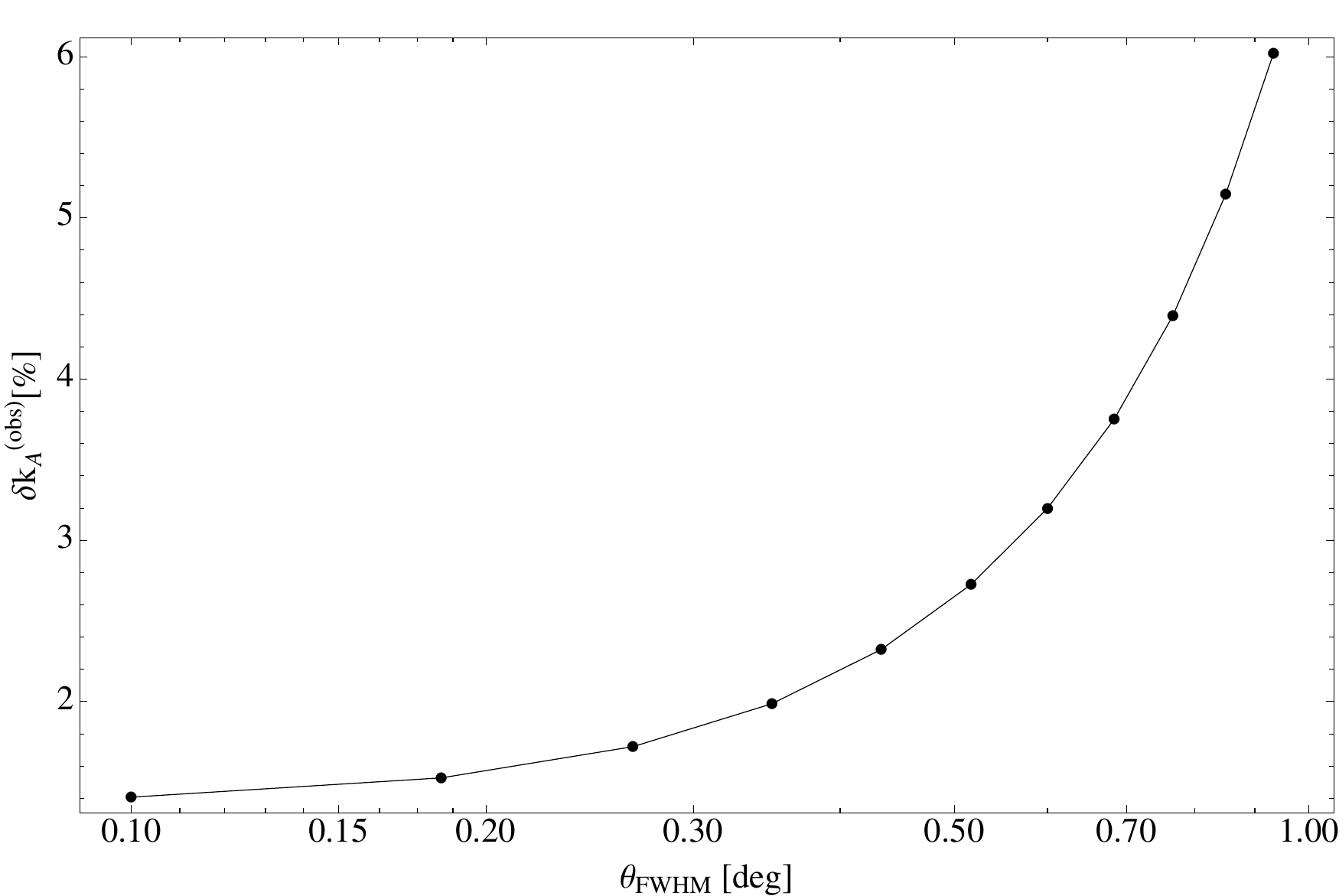}
\includegraphics[scale=0.4]{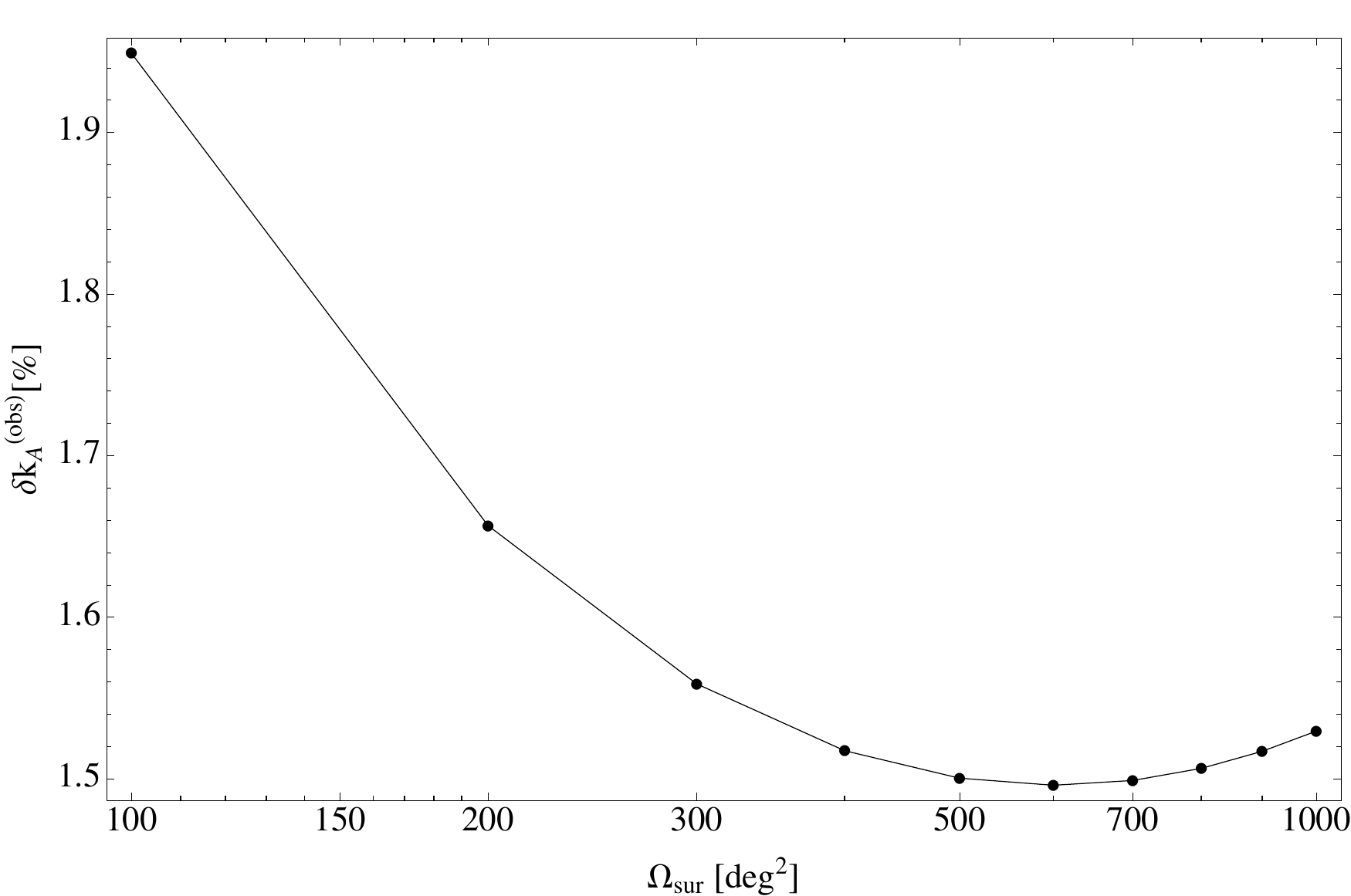}
\caption{Optimization of the measurement of $k_{\rm A}$ for case (B) - observations over the frequency range $600-900\,{\rm MHz}$. We have computed the projected error on the measurement of  $k_{\rm A}$ as a function of $\Delta\Omega_{\rm sur}$ and the beam size $\theta_{\rm FWHM}$.  In the top panel the contours are lines of constant $\delta k_{\rm A}$. The numbers on the contours are defined in the same way as in Fig.~\ref{fig:ratio3D_A}.  In the middle panel we present a 1D slice as a function of  $\theta_{\rm FWHM}$ with $\Delta\Omega_{\rm sur}=500\,{\rm deg}^2$ and in the bottom panel a 1D slice as a function of $\Omega_{\rm sur}$ with $\theta_{\rm FWHM}=10\,{\rm arc min}$.}
\label{fig:ratio3D_B}
\end{figure}

\section{Optimizing the performance and science potential of a single dish experiment}
\label{sec:optimise}

\noindent In this section we analyse the expected performance of a
single  dish telescope in two frequency ranges either side of the mobile phone band around $900-960\,{\rm MHz}$: (A) $f=960-1260\,{\rm MHz}$ $(0.13<z<0.48)$ and (B) $f=600-900\,{\rm MHz}$ $(0.58<z<1.37)$. We will compute projected errorbars on the
resulting power spectra  based on the methods presented in \citet{Seo:2010} using the 3D power spectrum (\ref{fig:PHI3Dzeq05}). In each case, (A) and (B), we will optimize the area covered, $\Omega_{\rm sur}$, and the resolution of the telecope, $\theta_{\rm FWHM}\propto 1/D$, where $D$ is the illuminated area of the dish, in order to give the best possible measurement of the acoustic scale. 

The standard expression for the projected error on a power spectrum measurement averaged over
a radial bin in $k$-space of width $\Delta k$ is \citep{Feldman:1994, Seo:2010}  
\be
\label{eq:error3D}
{\sigma_{P}\over P}=
\sqrt{2\frac{(2\pi)^3}{V_{\rm{sur}}}\frac{1}{4\pi k^2 \Delta k}}
\left(1+{\sigma^2_{\rm{pix}}V_{\rm{pix}}\over [{\bar T}(z)]^2W(k)^2P}\right),
\ee
where we have neglected the contribution from shot noise.  In order for the analysis to be simplified, it is necessary for us to make the errorbars on these so-called bandpower measurements uncorrelated. This requires that  $\Delta k>\pi/(r(z)\theta_{\rm min})$ where we assume that the observed sky area is a square patch with $\Omega_{\rm sur}=\theta_{\rm min}\times\theta_{\rm max}$ and $\theta_{\rm max}>\theta_{\rm min}$. In order to be able to detect the acoustic scale we need $\Delta k$ to be  less than a few times $k_{\rm A}$ and this can impose important limitations on the design of drift scan surveys where typically $\theta_{\rm max}\gg \theta_{\rm min}$.  $V_{\rm sur}$ is the survey volume given by
\be
\label{eq:vol}
V_{\rm{sur}}=\Omega_{\rm sur}\int_{z_{\rm min}}^{z_{\rm max}} dz \frac{dV}{dz d\Omega}\,,
\ee
where 
\be
\frac{dV}{dz d\Omega}=\frac{c r(z)^2}{H_0E(z)}\,.
\ee

The window function $W(k)$ models the instrument's angular  and frequency response.
Because our frequency resolution is relatively high, we can ignore
the instrument response function in the radial direction.
However, we have a finite angular resolution, which imposes a window function
\be
W(k)= \exp\left[-{1\over 2}k^2r(z)^2\left(\frac{\theta_{\rm{FWHM}}}
{\sqrt{8\rm{ln}2}}\right)^2\right].
\ee
The beam area is given by  $\Omega_{\rm pix}\propto \theta_{\rm FWHM}^2$ and the pixel volume is given by
\be 
V_{\rm pix}=\Omega_{\rm pix}\int_{z-\Delta z/2}^{z+\Delta z/2}dz{dV\over dz d\Omega}\,,
\ee
where $\Delta z$ is the redshift range corresponding to $\Delta f$, the nominal frequency channel width, which we will take to be $1\,{\rm MHz}$. 

\begin{figure}
\centering
\includegraphics[scale=0.4]{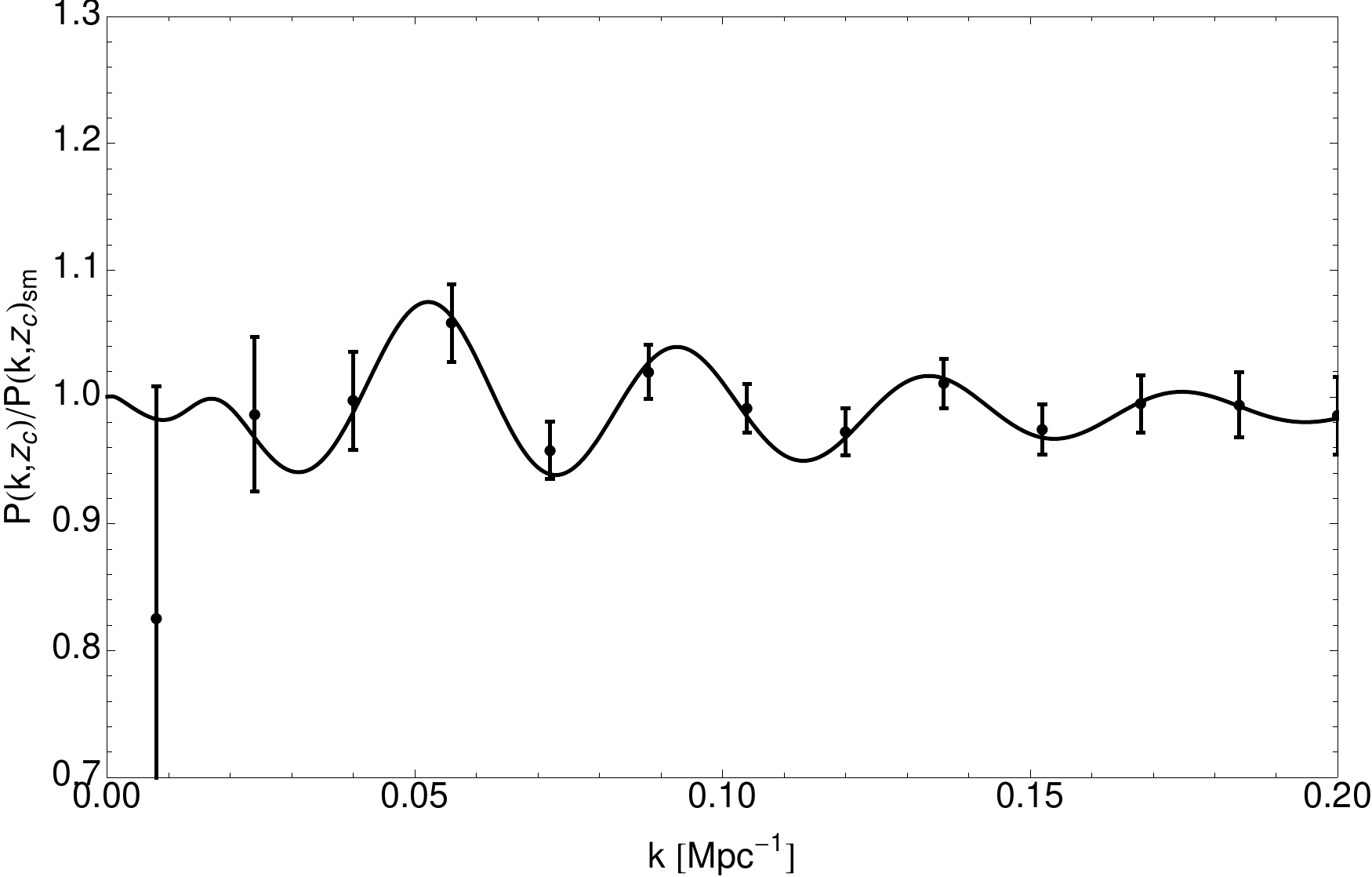}
\caption{Projected errors on the power spectrum (divided by a smooth power spectrum) expected for the survey described by the parameters used in the text and Table~\ref{tab:optimal} which is close the optimal value for Fig.\ref{fig:ratio3D_A}, that is, case (A) from Table \ref{tab:optimal}. We have used $\Delta k = 0.016\,{\rm Mpc}^{-1}$. The projected errors would lead to a measurement of the acoustic scale with a percentage fractional error of $2.4\%$. Note that the points do not lie on the line since they are band powers and hence are integrals over neighbouring values of $k$.}
\label{fig:spectrum_A}
\end{figure}

\begin{figure}
\centering
\includegraphics[scale=0.4]{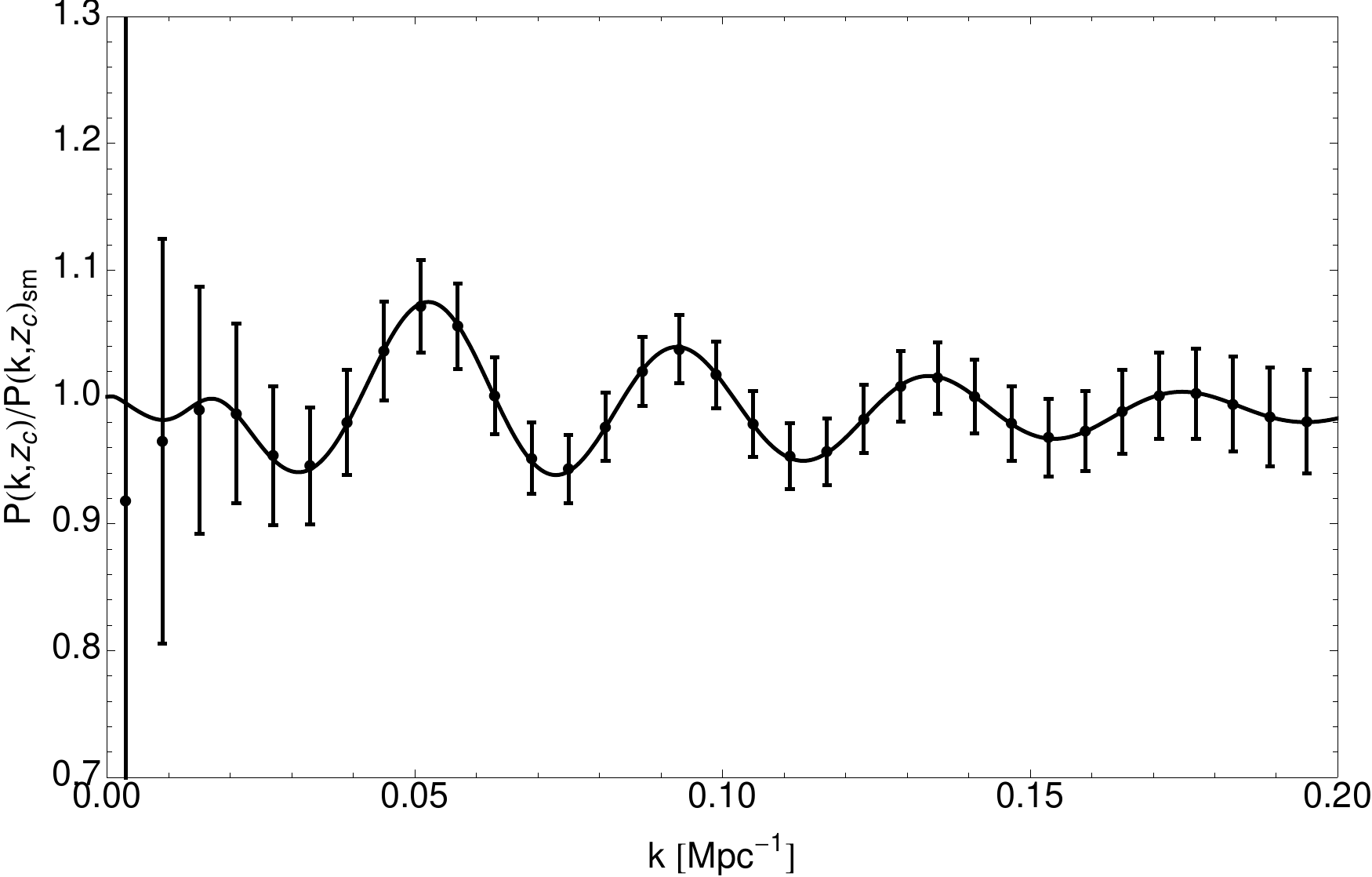}
\caption{Projected errors on the power spectrum (divided by a smooth power spectrum) expected for the survey described by the parameters used in the text and table~\ref{tab:optimal} which is close the optimal value for Fig.\ref{fig:ratio3D_B}, that is case (B) from table \ref{tab:optimal}. We have used $\Delta k = 0.006\,{\rm Mpc}^{-1}$. The projected errors would lead to a measurement of the acoustic scale with a percentage fractional error of $1.5\%$.}
\label{fig:spectrum_B}
\end{figure}

The pixel noise for a perfect, dual-polarisation, switch beam receiver measuring is given by 
\be 
\sigma_{\rm pix}={T_{\rm sys}\over \sqrt{t_{\rm pix}\Delta f}}\,,
\ee
where $T_{\rm sys}$ is the overall system temperature including contributions from the sky, ground and receiver, which we will assume to be $50\,{\rm K}$ - a value that is conservative for a well designed system based on room temperature low noise amplifiers (LNAs) - and $t_{\rm pix}$ is the time spent observing one pixel. This is related to the total observing time, $t_{\rm obs}$, by 
\be 
t_{\rm obs}={\Omega_{\rm sur}\over n_{\rm F}\Omega_{\rm pix}}t_{\rm pix}\,,
\ee
where $n_{\rm F}$ is the number of feedhorns at the focus of the telescope. In what follows we will keep $t_{\rm obs}$ and $n_{\rm F}$  fixed to be 1 year, that is a full year of on-source integration, something which may take around 2 years of elapse time to achieve, and 50, respectively.
In order to estimate the error on the measurement of $k_{\rm A}$ we will follow \cite{Blake:2003rh} and fit a decaying sinusoidal function to the projected data
\be
\label{eq:sinusoidal}
\frac{P(k)}{P_{\rm{ref}}}=1+Ak\exp\left[-\left(\frac{k}{0.1 \, h \, \rm{Mpc}^{-1}}\right)^{1.4}\right]
\sin \left(\frac{2\pi k}{k_{\rm A}}\right).
\ee
where the two parameters are  the acoustic scale, $k_{\rm A}$, and the  overall amplitude, $A$, which we will ultimately marginalize over.  

The parameter $k_{\rm A}$ is a function of the cosmological parameters and in particular the matter density measured relative to the critical density, $\Omega_{\rm m}$, and the equation of state parameter of the dark energy, $w=P_{\rm de}/\rho_{\rm de}=w_0+w_1z$. The angle-averaged power spectrum allows one to measure the isotropic BAO scale $k_{\rm A}=(k^2_\perp k_\parallel)^{1/3}$ in terms of the wavenumbers parallel to, $k_\parallel$, and perpendicular to, $k_{\perp}$, the line-of-sight. An efficient and quick method to do this is to use the approximations~\citep{Seo:2003,Glazebrook:2005mb}
\be
\label{kparkperp}
k_\parallel =\frac{H(z)}{H^{\rm{ref}}(z)}k^{\rm{ref}}_\parallel \;\;, 
k_\perp = \frac{d_{\rm A}^{\rm{ref}}(z)}{d_{\rm A}(z)}k^{\rm{ref}}_\perp\,,
\ee
in terms of reference values denoted by "ref" used for normalization purposes. Our reference cosmology is our input cosmology, with $w=-1=\rm{const}$, $\Omega_{\rm m}=0.27$ and
$h=0.71$.  This allows us to define the angle-averaged distance to redshift $z$ is given by \citep{Beutler:2011}
\be
d_{\rm V}(z)=\left(d_{\rm A}(z)^2{cz(1+z)^2\over H(z)}\right)^{1/3}\,,
\ee
which takes into account the angular diameter distance, $d_{\rm A}(z)$, and the radial distance $c/H_0E(z)$.

In Figs.~\ref{fig:ratio3D_A} and \ref{fig:ratio3D_B} we present projected percentage errors on the acoustic scale, $100\delta k_{\rm A}/k_{\rm A}$ as a function of the survey area, $\Omega_{\rm sur}$, and the beam size, $\theta_{\rm FWHM}$ for the redshift ranges given by cases A and B. First it is clear that in each case there are a wide range of survey parameters for which accurate measurements of $k_{\rm A}$ can be made. It will always be the case that reducing $\theta_{\rm FWHM}$ will lead to a better measurement of $k_{\rm A}$. However, beyond some point one gets into a situation where there is a very significant extra cost for very little extra return.  

In case A for $\Omega_{\rm sur}>2000\,{\rm deg}^2$ the contours of  constant $\delta k_{\rm A}/k_{\rm A}$ are almost vertical. This implies that for a fixed value of $\theta_{\rm FWHM}$, so long as one can achieve $\Omega_{\rm sur}>2000\,{\rm deg}^2$, optimisation of the survey area can only yield minor improvements. This is illustrated in the bottom panel of Fig.~\ref{fig:ratio3D_A}, where we fixed $\theta_{\rm FWHM}$. The optimum $\Omega_{\rm sur}$ is between 5000 and $6000\,{\rm deg}^2$ where $\delta k_{\rm A}/k_{\rm A}\approx 0.0225$, but for example if $\Omega_{\rm sur}=2000\,{\rm deg}^2$, as we will suggest for the BINGO instrument discussed in Section~\ref{sec:BINGO}, $\delta k_{\rm A}/k_{\rm A}\approx 0.024$, a difference which is within the uncertainties of the present calculations. The exact position of the optimum is governed by a balance between cosmic variance and thermal noise terms in equation\,(\ref{eq:error3D}); typically, it is where the two are approximately equal. In the middle panel of Fig.~\ref{fig:ratio3D_A} we fix $\Omega_{\rm sur}=2000\,{\rm deg}^2$ and vary only $\theta_{\rm FWHM}$. There is a significant improvement in $\delta k_{\rm A}/k_{\rm A}$ as one goes from $\theta_{\rm FWHM}=2\,{\rm deg}$ to around $40\,{\rm arcmin}$ but beyond that point there is little improvement. The reason for this is made apparent in Fig.~\ref{fig:spectrum_A} where we present the projected errors on $P(k)/P_{\rm smooth}(k)$, where $P_{\rm smooth}(k)$ is a smooth power spectrum chosen to highlight the BAOs. We see that as one goes to higher values of $k$, corresponding to higher angular resolution and lower values of $\theta_{\rm FWHM}$, the overall amplitude of the BAO signal decreases and therefore it is clear that the discriminatory power, in terms of measuring BAOs\footnote{We note that this might be different if we had chosen another science goal such as the measurement of neutrino masses.} comes from the first three oscillations in the power spectrum. Although measuring beyond this point will improve the constraint, the improvement does not justify the substantial extra cost involved; going from $\theta_{\rm FWHM}=40\,{\rm arcmin}$ to $20\,{\rm arcmin}$ would require the telescope diameter to increase by a factor of 2 and the cost by a factor of somewhere between 5 and 10. For the subsequent discussion of case A, we will concentrate on  $\theta_{\rm FWHM}=40\,{\rm arcmin}$ and $\Omega_{\rm sur}=2000\,{\rm deg}^2$ and the relevant survey parameters are presented in Table~\ref{tab:optimal} . This is not exactly the optimum point in the $\theta_{\rm FWHM}-\Omega_{\rm sur}$ plane, but it is sufficiently close to the optimum. $\theta_{\rm FWHM}=40\,{\rm arc min}$ can be achieved with an illuminated diameter of around $25\,{\rm m}$ at the relevant frequencies and $\Omega_{\rm sur}=2000\,{\rm deg}^2$ can be achieved with a drift scan strip of around $10\,{\rm deg}$ at moderate latitudes which is compatible with our choice of $\Delta k$. We note that the precise values of the optimum depend on the overall survey speed which is governed by $t_{\rm obs}n_{\rm F}/T_{\rm sys}^2$. Typically an increase in the overall survey speed will move the optimum survey area to a larger value.

\small
\begin{table*}
\caption{Table of the survey parameters used in Figs.~\ref{fig:spectrum_A} and \ref{fig:spectrum_B}. In both cases we have  chosen the values of $\theta_{\rm FWHM}$ and $\Omega_{\rm sur}$ so as to be close the  optimal values as described in the text. Included also are the projected measurements of the acoustic scale and estimates of the error $w$ (assumed constant) when all other cosmological parameters are known.\label{tab:optimal}}
\small
  \begin{tabular}{lccl}
\hline
\hline
Property & Low redshift (A) & High redshift (B)       \\ 
\hline
$z_{\rm centre}$ & 0.28   & 0.9  \\
$r(z_{\rm centre})$& $1100\,{\rm Mpc}$ & $3000\,{\rm Mpc}$ \\
$\Omega_{\rm sur}$& $2000\,{\rm deg}^2$ & $500\,{\rm deg}^2$ \\
$V_{\rm sur}$ & $1.2\,{\rm Gpc}^3$ & $3.1\,{\rm Gpc}^3$ \\
$\theta_{\rm FWHM}$& $40\,{\rm arcmin}$ & $10\,{\rm arcmin}$\\ 
$V_{\rm pix}$ & $810\,{\rm Mpc}^3$ & $590\,{\rm Mpc}^3$ \\
r.m.s. noise level ($\Delta f=1\,{\rm MHz}$) & $85\,\mu{\rm K}$ & $170\,\mu{\rm K}$ \\
\hline
$\delta k_{\rm A}/k_{\rm A}$ & 0.024 & 0.015 \\
$\delta w / w$ & 0.16 & 0.07 \\
\hline
\hline
 \end{tabular}
\end{table*}
\normalsize

\begin{figure}
\centering
\includegraphics[scale=0.6]{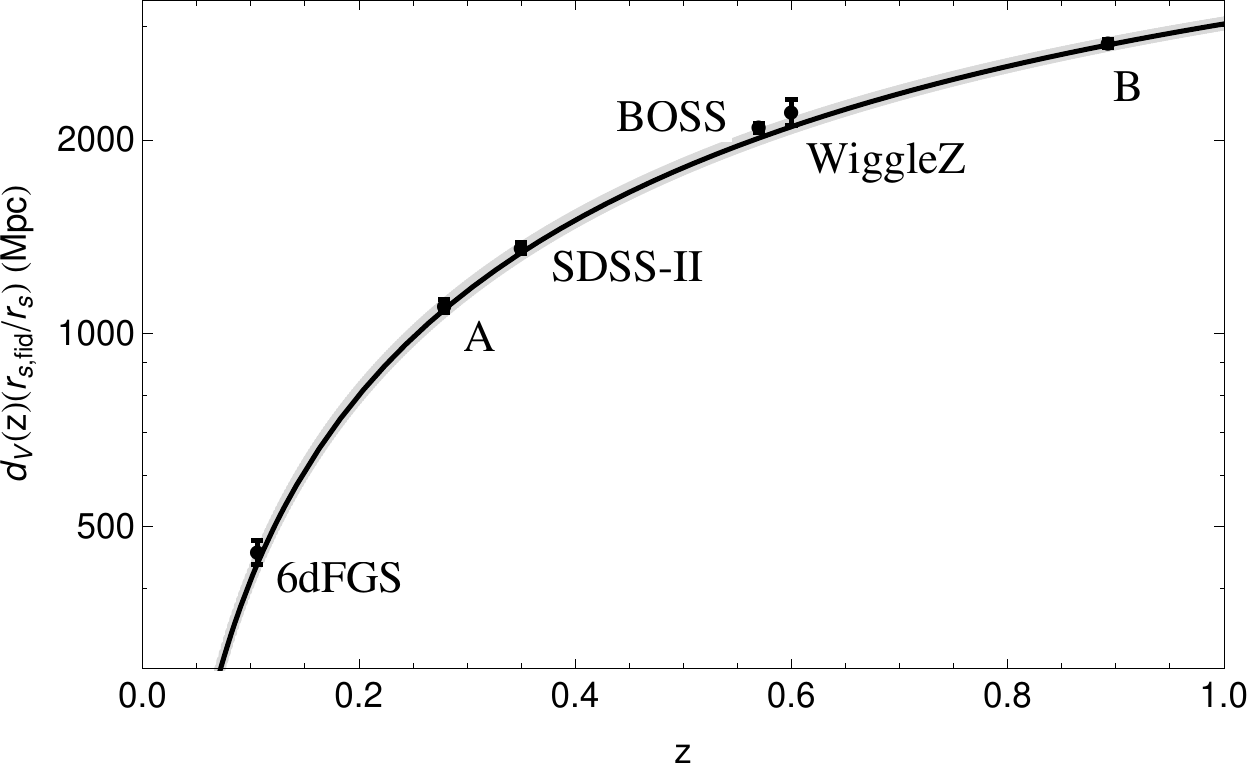}
\includegraphics[scale=0.6]{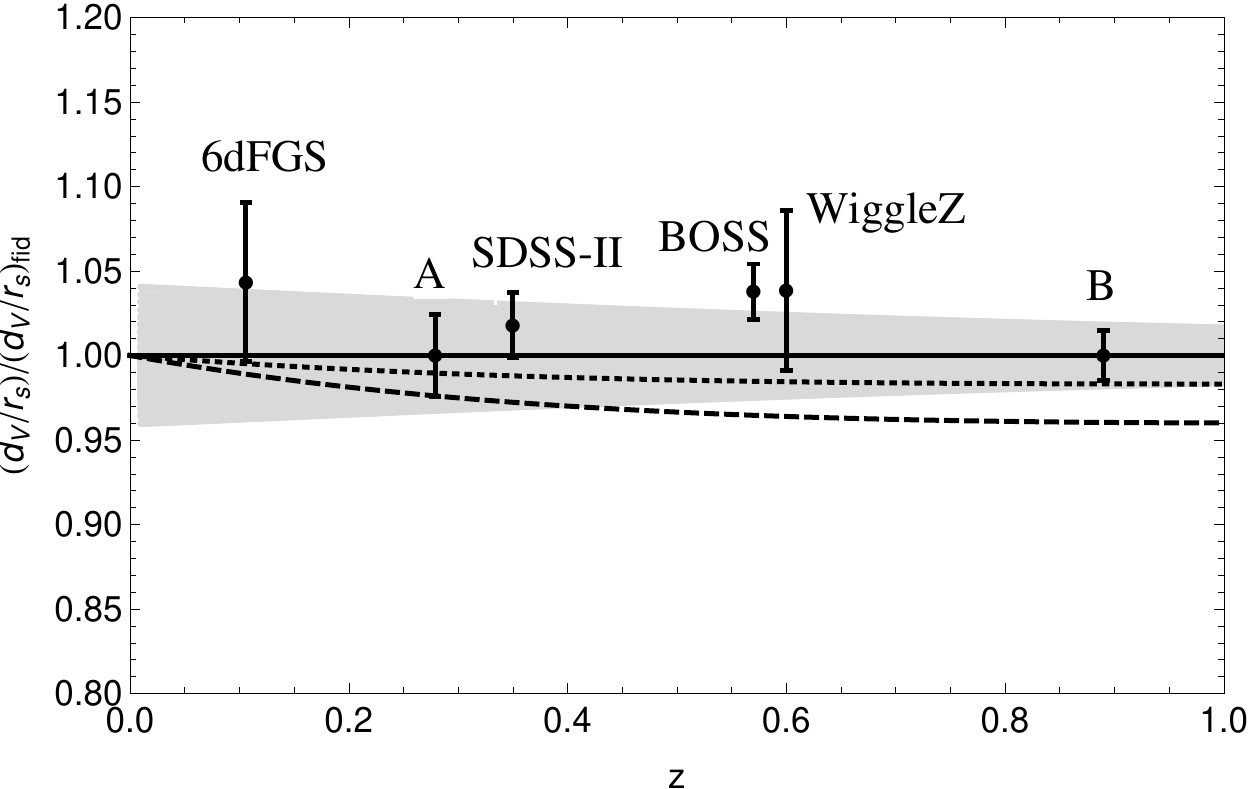}
\caption{Projected constraints on the Hubble diagram for the volume averaged distance, $d_{\rm V}(z)$, for cases A and B described by the parameters in table~\ref{tab:optimal}. Included also are the actual measurements made by 6dF, SDSS-II, BOSS and WiggleZ. The top panel is the absolute value and the bottom panel are the residuals from the fiducial model. On the bottom panel the shaded region indicates the range of $d_{\rm V}$ allowed by the $1\sigma$ constraint $\Omega_{\rm m} h^2$ from WMAP7 \citep{Komatsu:2011}. The dotted line is the prediction for $w_0=-0.84$ (which could in principle be ruled out by case A) and the dashed line is $w_0=-0.93$ (by case B).}
\label{fig:hubble}
\end{figure}

The situation for case (B) is qualitatively similar, but the exact values are somewhat different. Being centred at a much higher redshift, $z\approx 0.9$, means that the angular size of the BAO scale is much smaller and hence higher resolution is necessary to resolve the BAO scale and a smaller area needs to be covered to achieve the same balance between thermal noise and cosmic variance. We find that something close to optimal can be achieved for $\theta_{\rm FWHM}=10\,{\rm arcmin}$ and $\Omega_{\rm sur}=500\,{\rm deg}^2$. The projected errors for such a configuration are displayed in Fig. \ref{fig:spectrum_B} which would lead to $\delta k_{\rm A}/k_{\rm A}\approx 0.015$. In order to achieve this, one would need a single dish telescope with an illuminated diameter of $\sim 100\,{\rm m}$ which would have to be steerable, at least in some way. It would be relatively expensive to build a dedicated facility to perform such a survey, although, of course, such facilities do already exist, albeit with very significant claims on their time from other science programs. Perhaps an interferometric solution could be used to achieve this optimum: we find that an array of $\sim 250$  antennas with a diameter of $4\,{\rm m}$ spread over a region of diameter $\sim 100\,{\rm m}$, having the same value $T_{\rm sys}$, could achieve the same level of sensitivity. More generally an interferometer comprising $n_{\rm d}$ antennas of diameter $D$ spread over a region of diameter $d$ would be equivalent to a single dish of diameter $d$ with a focal plane array with $n_{\rm F}$ feedhorns, if $n_{\rm d}=d\sqrt{n_{\rm F}}/D$ so long as $n_{\rm F}<n_{\rm d}$, that is the filling factor of the interferometer is $<1$.

\begin{figure}
\centering
\includegraphics[scale=0.4]{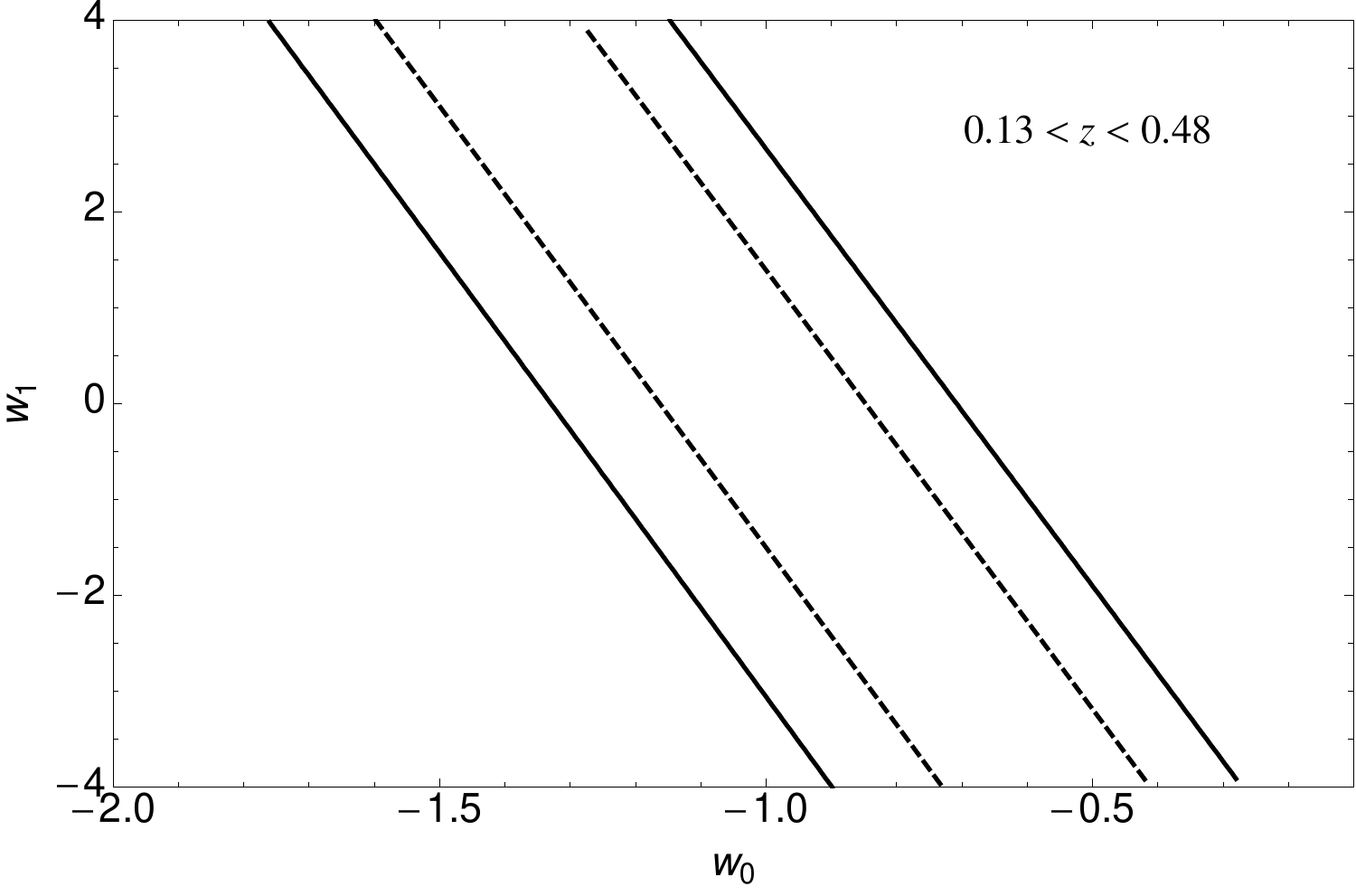}
\includegraphics[scale=0.4]{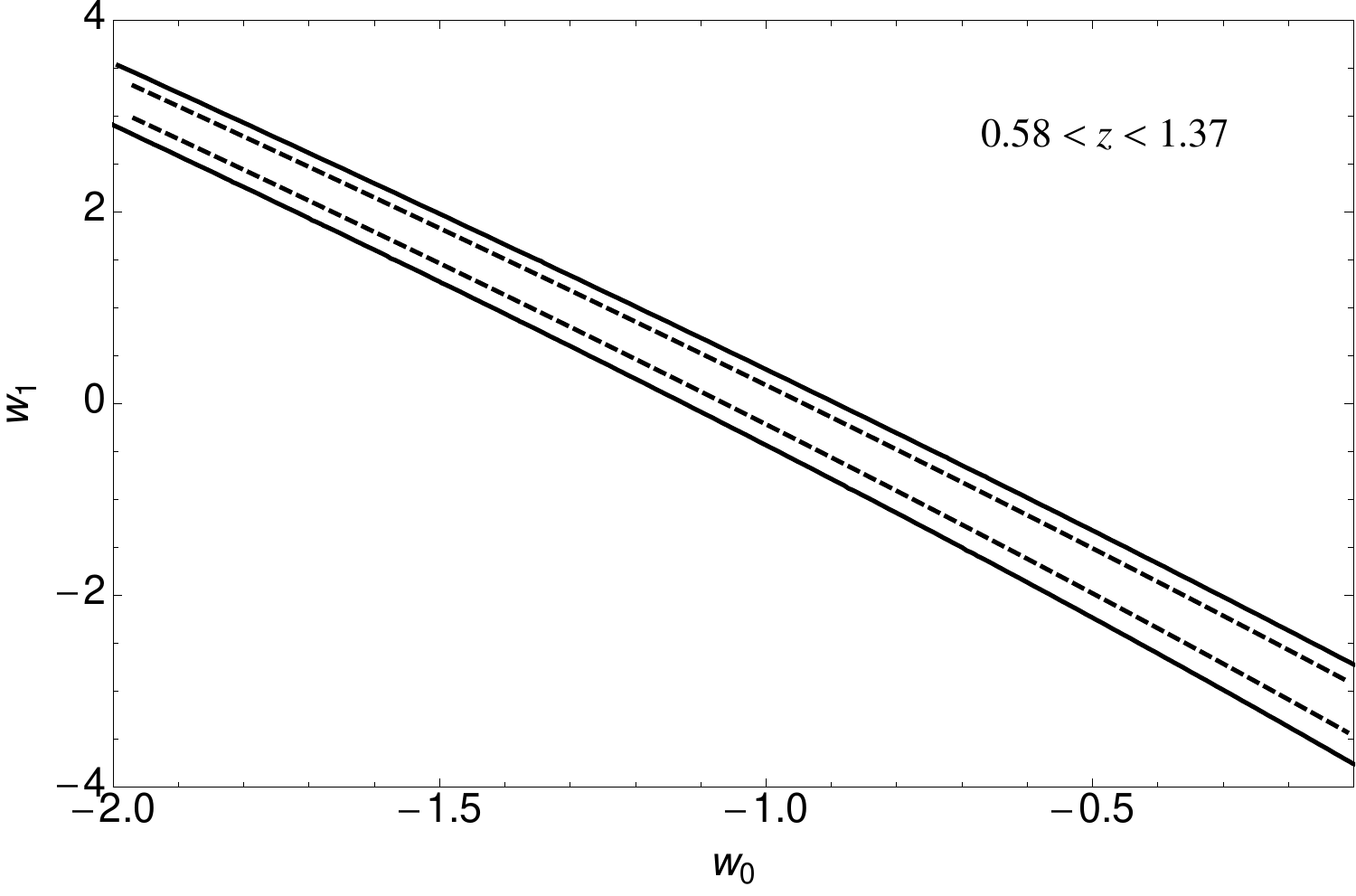}
\includegraphics[scale=0.4]{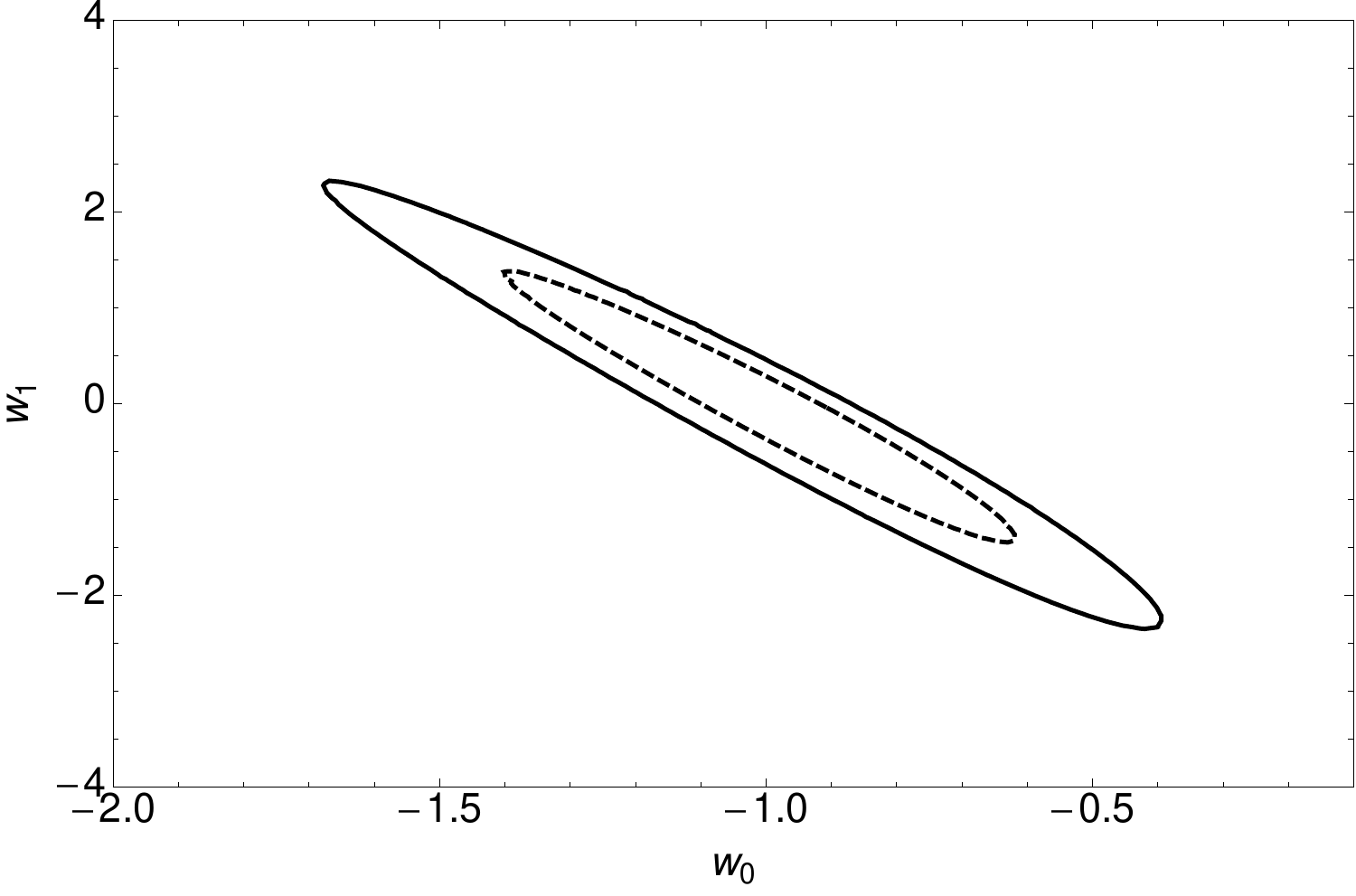}
\caption{Projected uncertainties on the dark energy parameters $w_0$ and $w_1$ for case A with $\theta_{\rm FWHM}=2/3\,{\rm deg}$ and $\Omega_{\rm sur}=2000\,{\rm deg}^2$ (top), case B with $\theta_{\rm FWHM}=10\,{\rm arcmin}$ and $\Omega_{\rm sur}=500\,{\rm deg}^2$ (middle) and a combination of the two (bottom) for the case where all other cosmological parameters are known. We have assumed that cases A and B lead to a measurement of the acoustic scale at the central redshift of the survey which means that there is an absolute parameter degeneracy between the two parameters, but the degenerate line is different at the two redshifts. Combination of the two removes the degeneracy.}
\label{fig:w0w1}
\end{figure}

In order to get an impression of the science reach of the proposed observations we have done two things. First, in Fig.~\ref{fig:hubble} we have plotted our projected constraints on the angle-averaged distance on the BAO Hubble diagram - both absolute and residuals - along with the state-of-the-art measurements from 6dFGS, SDSS-II, BOSS and WiggleZ. It is clear that the measurements that would be possible from instruments capable of achieving the design parameters of cases A and B would be competitive with the present crop of observations and those likely to be done in the next few years. They would also be complementary in terms of redshift range and add to the possibility of measuring the cosmic distance scale solely from BAOs. We note that it should be possible, in fact, to make measurements of the acoustic scale over the full range of redshifts probed by the intensity mapping surveys and it would be possible to place a number of data points on the Hubble diagram, albeit with larger errorbars. In order to compare with the present surveys we decided to present a single data point.

In addition, we have made simple estimates of the projected errors of the dark energy parameters assuming that all the other cosmological parameters, including $\Omega_{\rm m}$ are fixed. This will be a good approximation in the era of high precision CMB measurements made by, for example,  the {\it Planck} Satellite\footnote{We note that varying $\Omega_{\rm m}h^2$ modifies $d_{\rm V}(z)$ and the present limits on this parameter combination are included in Fig.~\ref{fig:hubble}. These will be substantially improved in the next couple of years.} . For constant $w$, the measurement of $\delta k_{\rm A}/k_{\rm A}=0.024$ possible in case A would lead to $\delta w/w=0.16$, whereas for case B, where $\delta k_{\rm A}/k_{\rm A}=0.015$ would lead to $\delta w/w=0.07$. For models with non-constant $w$, we have performed a simple likelihood analysis for the two-dimensional parameter space $w_0-w_1$ and the results are presented in Fig.~\ref{fig:w0w1} for cases A and B and a combination of the two. We see that since we have assumed a single redshift measurement of the acoustic scale, there is an exact degeneracy between the two parameters, but  the direction of the degeneracy is very different at the two different redshifts. When they are combined the degeneracy is broken and it is possible to estimate the two parameters separately. 

\section{Foreground contamination}

\label{sec:foregrounds}

\subsection{Overview}

To detect BAO in the faint HI signal, we will need to remove the
contributions from much brighter foreground emission. The observed
brightness temperature of the sky, $T_{\rm sky}(f,\bf{x})$ at a
frequency $f$ and position $\bf{x}$, consists of several
components
\begin{equation}
T_{\rm sky}(f,{\bf x})= T_{\rm CMB}(f,{\bf x}) + T_{\rm gal}(f,{\bf x}) + T_{\rm ps}(f,{\bf x}) + T_{\rm HI}(f,{\bf x})  ~,
\end{equation}
where $T_{\rm CMB}$ is the CMB temperature, $T_{\rm gal}$ is diffuse Galactic radiation, $T_{\rm ps}$ is emission from (mostly) unresolved extragalactic sources, and $T_{\rm HI}$ is the HI emission we are intending to detect. If the telescope and receiver are perfect, the total system temperature is given by $T_{\rm sys}=T_{\rm sky} + T_{\rm noise}$ i.e. no other contributions (e.g. from ground-spillover) will add to $T_{\rm sys}$.

The contribution from each foreground can be decomposed into a smooth
component, $\bar{T}(f)$,  and frequency and
position varying fluctuations, $\delta T (f,{\bf x})$, about this
smooth background
\begin{equation}
T(f,{\bf x}) = \bar{T}(f) + \delta T (f,{\bf x})\,.
\end{equation}

The foregrounds will be much brighter than the HI signal, by several
orders of magnitude. At $f=1$\,GHz, $\bar{T}_{\rm sky} \sim 5$\,K,
while the HI brightness temperature is $\sim 0.1$\,mK. However, our
objective is to measure fluctuations of the sky brightness temperature
as a function of both angular scale and redshift. Fluctuations of the
HI signal are expected to be at the same level as the smooth component
($\delta T_{\rm HI} \sim \bar{T}_{\rm HI} \sim 0.1$\,mK), while the
fluctuations of the foregrounds will typically be much less ($\delta T
\ll \bar{T}$). Furthermore, we can choose the coldest regions of sky
to minimise the foreground signal. We will show that regions
exist where the fluctuations in the  foregrounds are in fact a little less than two orders
of magnitude brighter than the HI signal. Given that the foregrounds
will be smooth in frequency, this gives us confidence that foreground
separation algorithms (e.g. \citep{Liu:2011}) can be employed that
will allow a detection of BAO.

Estimates of the foreground brightness temperature
fluctuations are presented  in the next sections for each component and are
summarised in Table~\ref{tab:fg}. 

\small
\begin{table*}
\caption{Summary of foregrounds for HI intensity mapping at 1\,GHz for an angular scale of $\sim 1^{\circ}$ ($\ell \sim 200$). The estimates are for a $10^{\circ}$-wide strip at declination $\delta=+45^{\circ}$ and for Galactic latitudes $|b|>30^{\circ}$. \label{tab:fg}}
\small
  \begin{tabular}{lccl}
\hline
Foreground             &$\bar{T}$   &$\delta T$               &Notes   \\
                       &[mK]        &[mK]                     &         \\
\hline
Synchrotron            &1700         &67                        &Power-law spectrum with $\beta \approx -2.7$.  \\
Free-free              &5.0         &0.25                        &Power-law spectrum with $\beta \approx -2.1$.    \\
Radio sources (Poisson)&  --        &5.5                         &Assuming removal of sources at $S>10$\,mJy.   \\
Radio sources (clustered) &--         &47.6                         &Assuming removal of sources at $S>10$\,mJy.   \\
Extragalactic sources (total)  &205    &48           &Combination of Poisson and clustered radio sources. \\
CMB           &2726        &0.07                       &Black-body spectrum, ($\beta=0$). \\
Thermal dust  &--            &$\sim 2\times 10^{-6}$                         &Model of \cite{Finkbeiner:1999}. \\
Spinning dust &--            &$\sim 2\times 10^{-3}$                         &\cite{Davies:2006} and CNM model of \cite{Draine98}. \\
RRL           &0.05            &$3\times 10^{-3}$                       &Hydrogen RRLs with $\Delta n = 1$.       \\
\hline
Total foregrounds        &$\sim 4600 $            &$\sim 82$    &Total contribution assuming the components are uncorrelated. \\
HI          &$\sim 0.1$                  &$\sim 0.1$         &Cosmological HI signal we are intending to detect. \\
\hline
\hline
 \end{tabular}
\end{table*}
\normalsize


\subsection{Galactic synchrotron radiation}
\label{sec:synchrotron}

Synchrotron radiation is emitted by electrons spiralling in the
Galactic magnetic field. The volume continuum emissivity is given by
the integration of the power radiated by an electron $P(f; E)$ at a
given energy $E$ and frequency $f$, and the cosmic ray energy
spectrum $N_{\rm e}(E)$
\begin{equation}
\epsilon_f = \frac{1}{4\pi} \int_{0}^{\infty} P(f; E) N_{\rm e}(E) dE.
\end{equation}
The integration over a distribution of electrons, which itself is smooth, results in
synchrotron spectra that are smooth, a property that we can exploit to
separate synchrotron emission from hydrogen line emission.
 
At frequencies $\sim 1$\,GHz observations indicate that
synchrotron emission is well-approximated by a power-law with spectral
indices in the range $-2.3$ to $-3.0$ \citep{Reich:1988}. The mean value $\bar{\beta} \approx -2.7$ and mean dispersion
$\sigma_\beta=0.12$ are measured by \citet{Platania:2003} in the range $0.4$--$2.3$\,GHz.

\begin{figure}
\centering
\vspace{-5mm}
\includegraphics[width=0.45\textwidth]{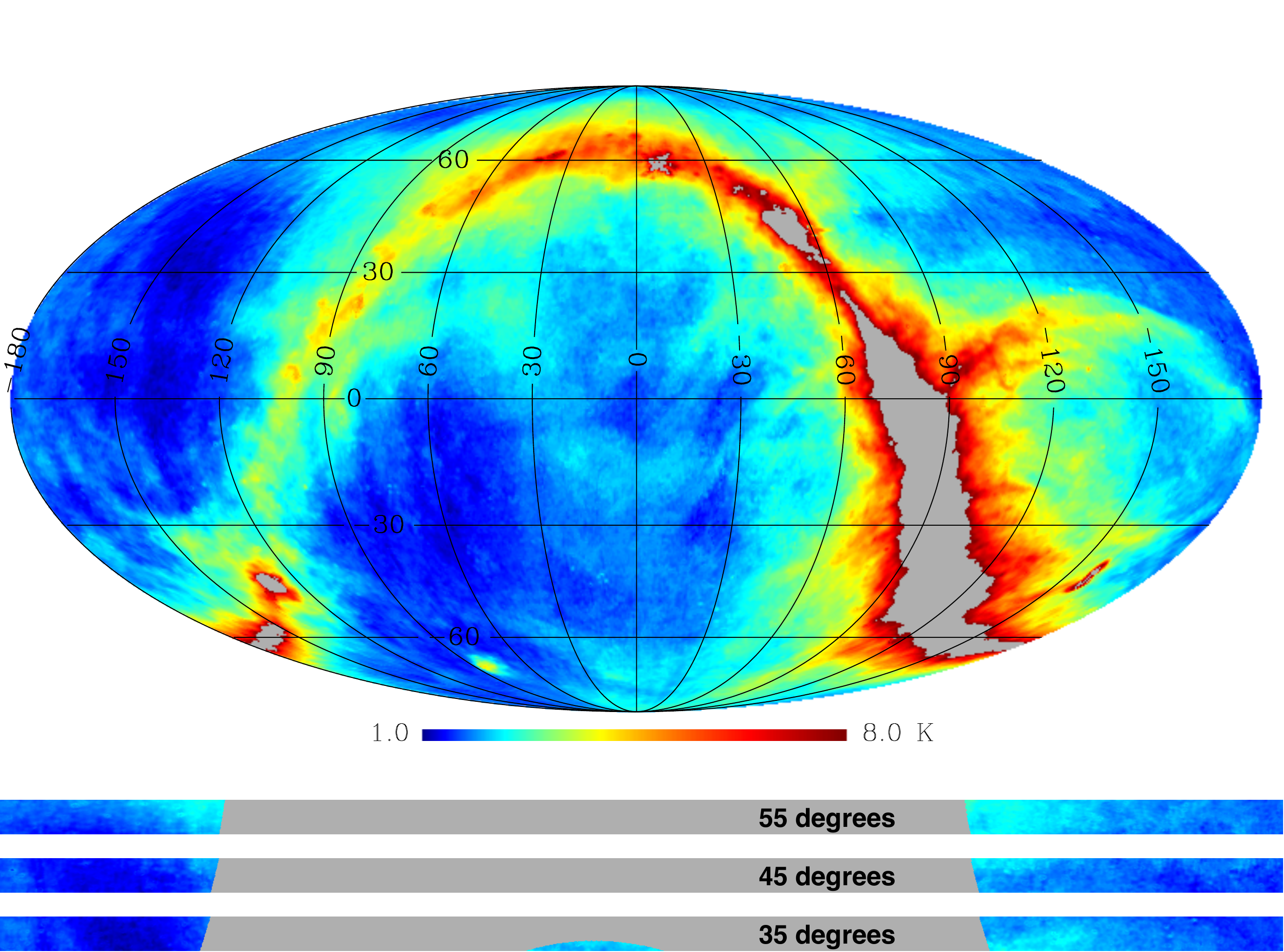}
\caption{In the top plot, we present the entire synchrotron sky in celestial (RA/Dec) coordinates, with RA$=0^{\circ}$ at the centre and increasing to the left. The map has been extrapolated from the all-sky 408\,MHz map \citep{Haslam:1982} to 1\,GHz using a spectral index $\beta=-2.7$ \citep{Platania:2003}. The colour-scale is asinh from 1\,K to 8\,K. The saturated pixels $(T>8$\,K) are shown as grey. In the bottom plot, we present three drift scan strips, each $10^{\circ}$ wide, located at declinations of $35^{\circ}$ (bottom), $45^{\circ}$ (middle) and $55^{\circ}$ (top).}
\label{fig:synchmaps}
\end{figure}

A number of large-area surveys are available at frequencies of $\sim
1$\,GHz, most notably the all-sky 408\,MHz map of
\citep{Haslam:1982}. We use the NCSA destriped and source-removed
version of this map\footnote{http://lambda.gsfc.nasa.gov/} to estimate
the contribution of synchrotron emission at 1\,GHz. The 408\,MHz map
is extrapolated to 1\,GHz using a spectral index of $\beta=-2.7$. The
map is shown in RA-Declination coordinates in
Fig.~\ref{fig:synchmaps}. The Galactic plane is evident as well as
large-scale features that emit over large areas of the high latitude
sky. The most prominent feature is the North Polar Spur which
stretches over $\sim 100^{\circ}$ of the northern sky. The mean
temperature at high Galactic latitudes $(|b|>30^{\circ})$ is $T_{\rm
sky}\approx 2$\,K with r.m.s. fluctuations in maps of $1^{\circ}$ resolution
of $\delta T_{\rm Gal}\approx 530$\,mK. The r.m.s. fluctuations are dominated
by contributions from the large-scale features.

Fortunately, there exist large areas of low synchrotron temperature in the north. The declination strip at $\delta \sim +40^{\circ}$ is well known to be low in foregrounds \citep{Davies:1996} and is accessible by observing sites in the north. Furthermore, we are interested in the fluctuations at a given angular scale, or multipole value $\ell$, rather than the r.m.s. value in the map at a given resolution. The r.m.s. estimate from the map will be considerably larger due to fluctuations on larger angular scales than the beam.

To estimate the foreground signal at a given $\ell$-value we calculated the power spectrum of the $\delta=+45^{\circ}$ strip after masking out the Galactic plane ($|b|>30^{\circ}$). This was achieved by using the {\sc Polspice}\footnote{{\tt http://www2.iap.fr/users/hivon/software/PolSpice/.}} software, which calculates the power spectrum from the 2-point pixel correlation function \citep{Szapudi:2001,Chon:2004}. We apodized the correlation function with a Gaussian of width $\sigma=10^{\circ}$ to reduce ringing in the spectrum at the expense of smoothing the spectrum. A power-law fit was then made to estimate the foreground signal at $\ell=200$ ($\sim 1^{\circ}$). The result is  given in Table~\ref{tab:fg}.

For a $10^{\circ}$ wide strip centred at $\delta=45^{\circ}$ we find that the $\bar{T}_{\rm Gal}=1.7$\,K with fluctuations of $\delta T_{\rm Gal}=67$\,mK. Utilising these cool regions will be important for detecting the faint HI signal. The exact choice of strip will be a critical factor to be decided before construction of a fixed elevation telescope while the Galactic mask to be applied can be chosen during the data analysis step.

\subsection{Free-free radiation}
\label{sec:freefree}

Free-free radiation (thermal bremsstrahlung) is emitted by electrons interacting with ions. At radio frequencies, this comes from warm ionized gas with typical temperatures of $T_e \sim 10^4$\,K. Away from the Galactic plane, the radiation is optically thin and its brightness temperature is  given by
\begin{eqnarray}
T_{\rm ff} \approx &90\,{\rm mK}\left({T_{\rm e}\over {\rm K}}\right)^{-0.35}\left({f\over {\rm GHz}}\right)^{-2.1}\left({\rm EM}\over {\rm cm}^{-6}\,{\rm pc}\right)&\,,
\end{eqnarray}
where EM is the emission measure representing the integral of the
electron density squared along the line-of-sight (EM$=\int n_{\rm e}^2
dl$).  The free-free spectrum is therefore a well-defined power-law
with a temperature spectral index $\beta=-2.1$ which acts to flatten
the spectral index of the total continuum where it has a brightness temperature
comparable to that of the synchrotron emission.

The EM can be estimated using H$\alpha$ measurements, which can then be converted to radio brightness temperature assuming $T_{\rm e}$ is known and the dust absorption is small \citep{Dickinson:2003}. For $|b|>30^{\circ}$, where $T_{\rm e}\approx 8000$\,K, EM$\,\sim 1.3$\,cm$^{-6}$\,pc at 1\,GHz, this corresponds to $\bar{T}_{\rm ff} \sim 5$\,mK. For the $10^{\circ}$ wide strip at $\delta=+45^{\circ}$ the free-free brightness temperature fluctuations are $\delta T_{\rm ff}=0.25$\,mK at $\ell=200$. The free-free component is therefore considerably weaker than the synchrotron component across most of the sky, but will act to (slightly) flatten the spectrum at higher frequencies.


\subsection{Extragalactic foregrounds}
\label{sec:extragalactic}

Extragalactic radio sources are an inhomogeneous mix of radio galaxies,
quasars and other objects. These contribute an unresolved 
foreground\footnote{For convenience we refer to extragalactic sources
as a foreground though both lower redshift (foreground)  and higher redshift (background) sources will contaminate the HI signal.}. The sources are often
split into two populations based on their flux density spectral index,
$\alpha$ ($S \propto \nu^{\alpha}$): steep spectrum sources with
$\alpha < -0.5$ and flat spectrum sources with $\alpha>-0.5$. At low
frequencies ($\lesssim 1$\,GHz), steep spectrum sources with $\alpha
\approx -0.75$ will dominate \citep{Kellermann:1968}.

\begin{figure}
\includegraphics[width=0.45\textwidth]{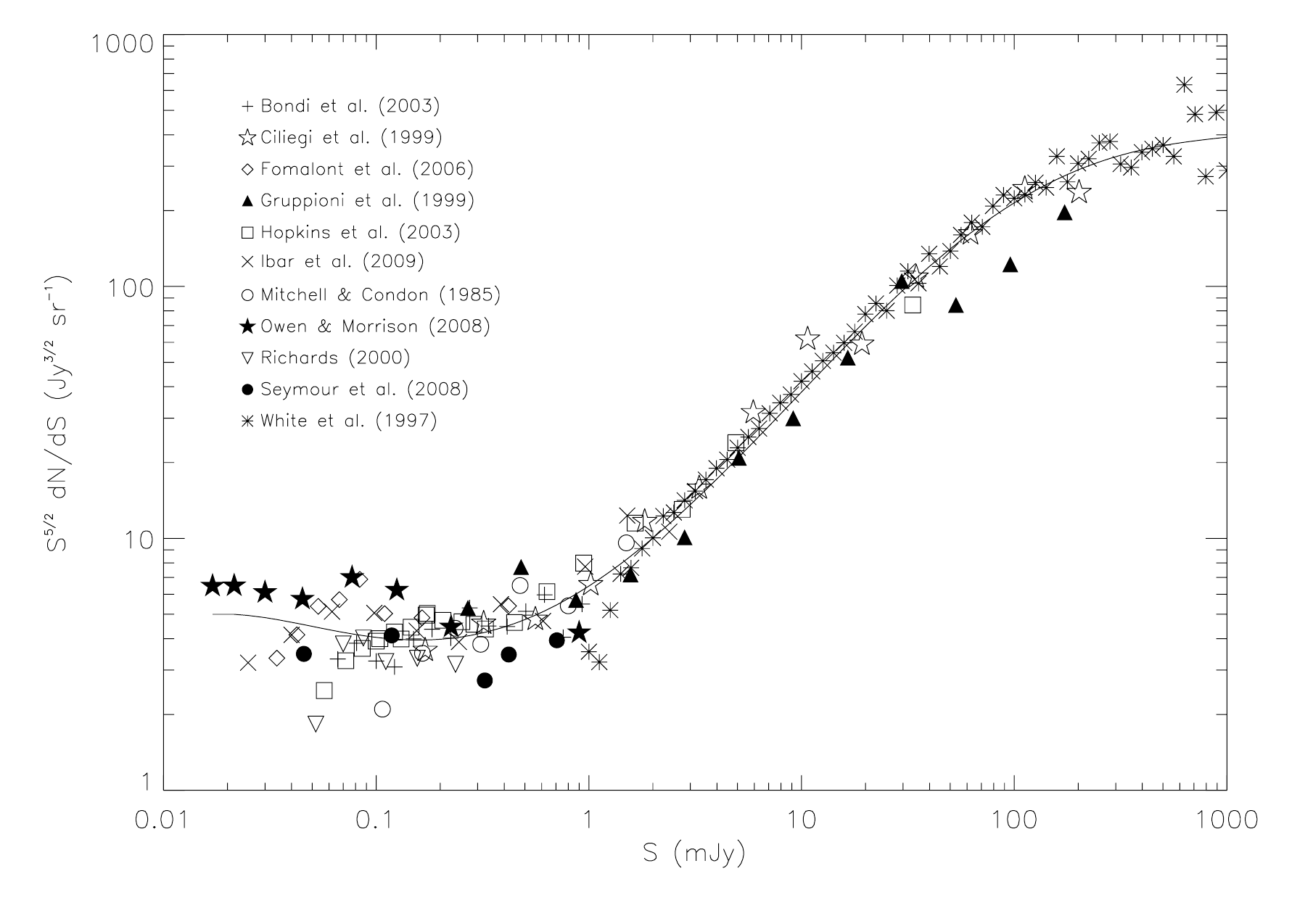}
\caption{The compilation of differential radio sources counts at 1.4\,GHz which we have used. The solid line is the best-fitting 5th order polynomial.}
\label{fig:sourcecount}
\end{figure}

The contribution of point sources, $T_{\rm ps}$, can be calculated from the differential source count, $dN/dS$, representing the number of sources per steradian, $N$, per unit flux, $S$. A number of compilations of source counts are available. We chose to use data collected from continuum surveys at 1.4\,GHz data between 1985 and 2009, summarised in Table~\ref{tab:source_surveys}. Fig.~\ref{fig:sourcecount} shows the Euclidean normalized differential source count for our compilation. We have fitted a 5th order polynomial to these data of the form
\begin{equation}
\textrm{log}\left(\frac{S^{2.5}dN/dS}{N_0} \right) = \sum_{i=0}^5 a_i \left[ \textrm{log}\left( \frac{S}{S_0} \right ) \right]^{i}~,
\end{equation}
where $N_0=1$\,Jy\,sr$^{-1}$ and $S_0=1$\,Jy are normalizing constants. A least-squares fit gives the best-fitting coefficients: $a_0=2.593$, $a_1=0.093$, $a_2=-0.0004$, $a_3=0.249$, $a_4=0.090$ and $a_5=0.009$. These values are compatible with those of \cite{Vernstrom:2011}.
\small

\begin{table}
\centering
\caption{References used for source surveys at $\sim 1.4$\,GHz used for compilation of source counts.\label{tab:source_surveys}}
\small
  \begin{tabular}{c}
\hline
Reference \\
\hline
\cite{Bondi:2003}     \\
\cite{Ciliegi:1999}     \\
\cite{Fomalont:2006}     \\
\cite{Gruppioni:1999}     \\
\cite{Hopkins:1999}     \\
\cite{Ibar:2009}     \\
\cite{Mitchell:1985}     \\
\cite{Owen:2008}     \\
\cite{Richards:2000}     \\
\cite{Seymour:2008}     \\
\cite{White:1997}     \\
\hline
\hline
 \end{tabular}
\end{table}
\normalsize

Assuming we can subtract all sources with flux density $S>S_{\rm max}$, we can calculate the mean temperature
\begin{equation}
\bar{T}_{\rm ps} =\left({dB\over dT}\right)^{-1} \int_{0}^{S_{\rm max}} S \frac{dN}{dS}dS\,,
\end{equation}
where $dB/dT=2kf^2/c^2$, $c$ is the speed of light and $k$ is the
Boltzmann constant. We find $\bar{T}_{\rm ps}=97$\,mK at 1.4\,GHz for
$S_{\rm max}=\infty$.  We would expect to be able to subtract the
brightest radio sources down to $S_{\rm max}=10$\,mJy using the NRAO
VLA Sky-Survey (NVSS) which is complete to $3.4$\,mJy
\citep{Condon:1998} although there will be a residual contribution due
to the variability of radio sources. Subtracting sources stronger than
10~mJy reduces the mean temperature of $\bar{T}_{\rm ps}=83$\,mK at
1.4\,GHz. Although this is a useful thing to do, we clearly see that the bulk
of the background temperature arising from extragalactic sources is
contributed by the weakest ones and this is precisely the point where the uncertainty in the counts is greatest. Extrapolating the course counts to 1\,GHz with
the mean temperature spectral index $\beta=-2.7$ gives $\bar{T}_{\rm
ps}=205$\,mK.

Before we continue, we note that there is evidence of an additional
component to the sky brightness as measured by the ARCADE experiment
at 3\,GHz \citep{Fixsen:2011}. They measured an excess component of
radiation with a brightness temperature of $\sim 1$\,K at 1\,GHz and
with a spectral index $\beta \approx -2.6$. Given the extreme
difficulty in making absolute sky brightness measurements further
observations, perhaps at lower frequency where the claimed excess is
stronger, are required to firmly establish its reality. However, if it
is real, this could require either a re-interpretation of Galactic
radiation contributions or be evidence for a new population of
extragalactic sources \citep{Seiffert:2011} - cf our point concerning the sensitivity of the computation of $T_{\rm ps}$ to the lower cut-off. For our initial analysis
presented here, we do not consider this result further although we
note that this could increase our foreground estimates significantly.

There are two contributions to $\delta T_{\rm ps}$. The first is due
to randomly (Poisson) distributed sources. However, if the sources
have a non-trivial two-point correlation function, then there is a
contribution due to clustering. We quantify the spatial fluctuations
in $\delta T_{\rm ps}$ in terms of the angular power spectrum,
$C_{\ell}=C_{\ell}^{\rm Poisson}+C_{\ell}^{\rm clustering}$, which is
the Legendre transform of the two point correlation function
$C(\theta)$.

Poisson distributed sources have a white power spectrum given by
\begin{equation}
C_{\ell}^{\rm Poisson} = \left({dB\over dT}\right)^{-2}\int_{0}^{S_{\rm max}} S^2 \frac{dN}{dS} dS~.
\end{equation} 
In terms of multipoles, $\ell$ (where $\theta \sim
180^{\circ}/\ell)$, the magnitude of the temperature fluctuations at
any given angular scale is given by $\delta T = \sqrt{\ell (\ell+1)
C_{\ell}/2\pi}$. Using our source count fit above to integrate up to
$S_{\rm max}=10$\,mJy and extrapolating to 1\,GHz leads to
fluctuations of $\delta T_{\rm ps}= 5.5$\,mK at $\ell=200$ ($1\,{\rm deg}$)

The clustered component of radio sources acts to increase pixel-pixel correlations. The power spectrum due the clustered sources can be simply estimated as $C_{\ell}^{\rm cluster}= w_{\ell} {\bar T}_{\rm ps}^2$, where $w_{\ell}$ is the Legendre transform of the angular correlation function, $w(\theta)$ \citep{Scott:1999}. The clustering of radio sources at  low flux densities ($<10$\,mJy) is not well-known. To make an estimate, we use $w(\theta)$ measured from NVSS, which can be approximated as $w(\theta)\approx (1.0 \pm 0.2)\times10^{-3} \theta^{-0.8}$ \citep{Overzier:2003}. Legendre transforming this gives $w_{\ell}\approx 4.8\times10^{-3} \ell^{-1.2}$. Although measured directly for sources at a higher flux density than those relevant here, the overall amplitude and power law are compatible with those expected if the galaxies follow clustering of cold dark matter.

\begin{figure}
\includegraphics[width=0.45\textwidth]{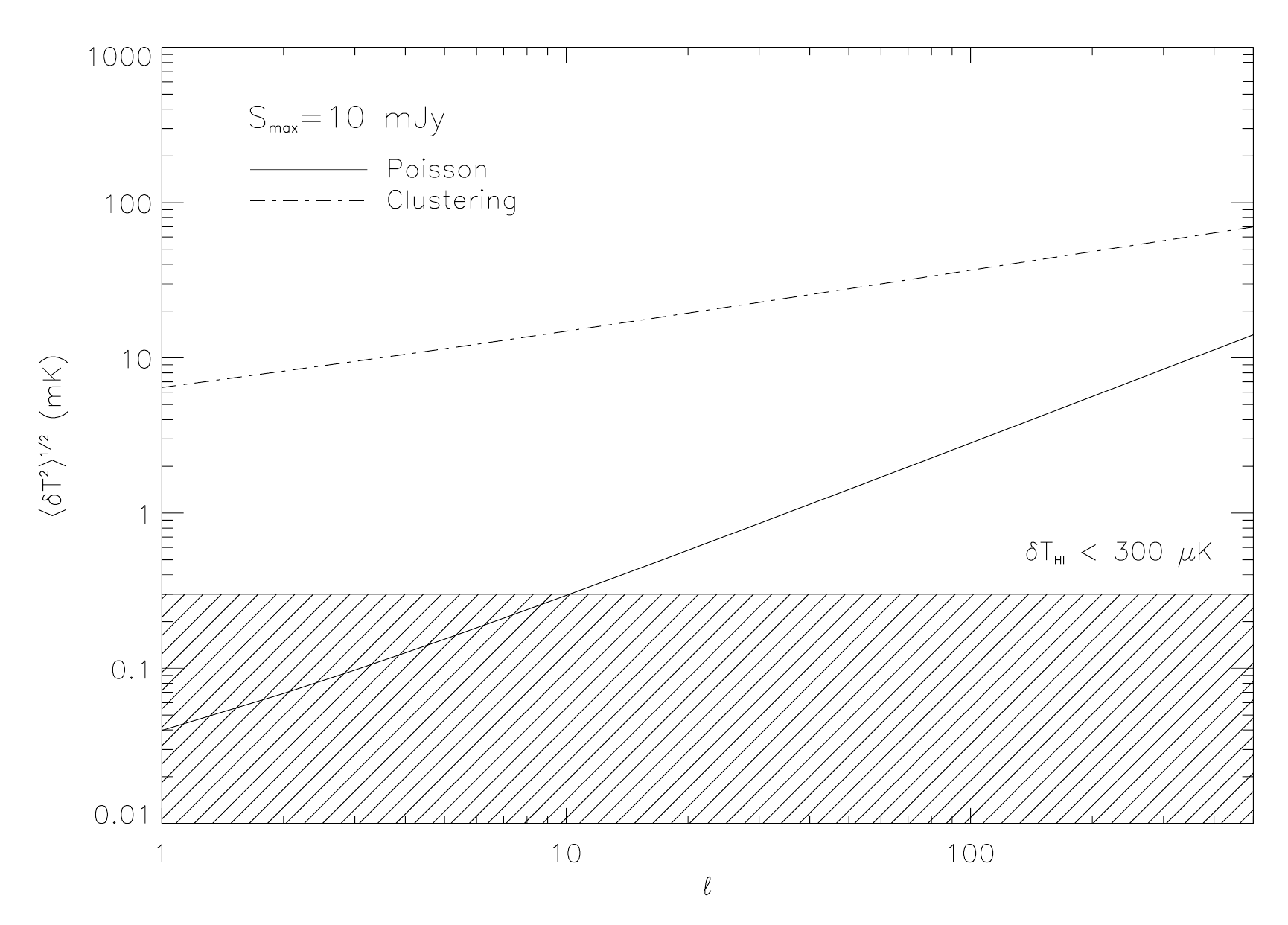}
\caption{The Poisson and clustering contribution to the sky brightness temperature fluctuations for a flux cut-off value of $S_{\rm max}=10$\,mJy. The brightness temperature fluctuations due to the 21\,cm signal is expected to be $\lesssim 300$\,$\mu$K at redshifts $z \approx 0.5$ and is indicated by the hatched region. \label{fig:sources_powerspectra}}
\end{figure}

The contribution to $\delta T_{\rm ps}$ at 1\,GHz on angular scales $\sim1^\circ$ from clustered sources, assuming a cut-off flux density of 10\,mJy, is $\delta T_{\rm ps}\approx 47.6$ mK. Plotting both the Poisson and clustering contributions to $\delta T_{\rm ps}$, it can be seen in Fig.~\ref{fig:sources_powerspectra} that the contribution from the clustering component appears to be dominant over the Poisson fluctuations for multipoles $\ell<500$ and is approximately an order of magnitude above the fluctuations in the HI signal.

The total contribution to $\delta T_{\rm ps}$ is then the quadrature
sum of the Poisson and clustered components (they are uncorrelated),
i.e. $C_{\ell}^{\rm ps} = C_{\ell}^{\rm Poisson} + C_{\ell}^{\rm
clustering}$ which leads to $\delta T_{\rm ps}=\sqrt{\delta T_{\rm
Poisson}^2 + \delta T_{\rm cluster}^2} \approx  48$\,mK. We note that there
is considerable uncertainty in this estimate, particularly with the
clustered component. The extrapolation of our source count into the
low flux density ($\ll 0.1$\,mJy) regime may be unreliable
(e.g. because the ARCADE results may imply an unexpectedly large
population of very weak sources) and can only be firmly established by
ultra-deep radio surveys with upcoming telescopes (e.g. \citet{Norris:2011}).


\subsection{Other foregrounds}

A number of other foregrounds could also contribute to $T_{\rm sky}$. Other
diffuse continuum emission mechanisms include thermal dust and
spinning dust emissions. Both of these are expected to be negligible
at $\nu \sim 1$\,GHz. For example, extrapolation of the thermal dust
model of \cite{Finkbeiner:1999}, at $1^{\circ}$ resolution predicts
$\delta T \sim 2 \times 10^{-6}$\,mK for $|b|>30^{\circ}$. Similarly,
extrapolating the anomalous microwave emission observed at 23\,GHz by
{\it WMAP} \citep{Davies:2006} and assuming it is all due to spinning
dust emission from the cold neutral medium (CNM; \citealt{Draine98}),
gives $\delta T \sim 2 \times 10^{-3}$\,mK and can therefore be safely neglected.

There are a number of spectral lines which emit in the radio band. The most
prevalent in the frequency range of interest are radio
recombination lines (RRLs), primarily from hydrogen. The frequencies
of these lines are given by the Rydberg equation, for example,
H187$\alpha$ is at a rest frequency of 997.63\,MHz. The intensity of
these lines can be estimated from the free-free continuum temperature,
$T_{\rm ff}$, using the following equation \citep{Rohlfs:2004}
\begin{equation}
\frac{\int T_{\rm L} dV}{T_{\rm ff}\Delta V} = 6.985 \times 10^3 \frac{1}{a(T_e)} \frac{1+n({\rm H})}{n({\rm He})}\bigg({T_{\rm e}\over {\rm K}}\bigg)^{-1.15}\bigg({\nu\over{\rm GHz}}\bigg)^{1.1}\,,
\end{equation}
where $a(T_e,\nu)$ is a correction factor, which at low frequencies is very close to 1 \citep{Mezger}. 
For a typical linewidth of 25\,km\,s$^{-1}$, assuming $T_{\rm e}=8000$\,K
and a 8\,\% helium to hydrogen ratio, the line temperature will be
$\sim 1$\,\% of the free-free continuum temperature at 1\,GHz. Given
the estimates for Galactic foreground emission from Section
\ref{sec:freefree}, we expect $\bar{T}_L \sim 0.05$\,mK and $\delta T_L \sim 2.5 \times 10^{-3}$\,mK. Therefore,
RRLs are not expected to be an important foreground either. Nevertheless, it would be wise to omit frequency channels that are known to contain the
brightest ($\Delta n=1$) hydrogen lines, which occur at intervals of
approximately $\Delta f \approx 20$\,MHz. We note, however, that
these recombination lines contain much astrophysically interesting
information about the Galaxy and therefore are likely to provide
significant spin-off science from any radio BAO project.

\section{The BINGO concept}
\label{sec:BINGO}

\noindent The discussion presented in Section~\ref{sec:optimise} suggests that in order to get
a good detection of BAO at low redshifts (case A from the optimisation study) a suitable single telescope should satisfy the following
design constraints:

\begin{itemize}

\item an angular resolution defined by $\theta_{\rm FWHM}\approx 40\,{\rm arcmin}$ corresponding to an under-illuminated primary reflector  with an edge taper $<-20\,{\rm dB}$;

\item operate in the frequency range above 960~MHz with a bandwidth that is as wide as possible  - exactly how wide and
exactly where the band is positioned will ultimately be determined by
those parts of the spectrum relatively free of radio frequency
interference (RFI);

\item frequency resolution should be $\sim$1~MHz or better;

\item frequency baseline should be as free as possible from
instrumental ripples, particularly standing waves \citep{Rohlfs:2004};

\item the receivers should be exceptionally stable in order to allow
thermal-noise-limited integrations of a total duration long enough to reach
a surface brightness limit $\sim100\mu$K per pixel;

\item the sky coverage as large as possible, but at the very
least $>$2000~deg$^2$ with a minimum size in the smallest dimension of $10\,{\rm deg}$;

\item the number of feeds should be $n_{\rm F}>50$ in order to achieve a sufficiently large signal-to-noise ratio;

\item sidelobe levels as low as possible so as to minimize
the pickup of RFI and emission from strong sources;

\item beam ellipticity $<0.1$ in order to allow map-making and power spectrum analysis routines to work efficiently;

\end{itemize}
For reasons of stability, simplicity, and cost, a telescope with no
moving parts would be very desirable. We suggest that the following
conceptual design could deliver what is required. We first concentrate
on the telescope and then on the receiver design. We have named the overall project BINGO which stands for {\bf B}AO from {\bf I}ntegrated {\bf N}eutral {\bf G}as {\bf O}bservations.

\subsection {The telescope and focal plane}

Operation at frequencies between $960\,{\rm MHz}$ and $1260\,{\rm MHz}$ is a practical possibility because it avoids strong RFI from mobile phone down-links which occupy the band up to 960 MHz. In order to achieve the desired resolution (better than $40\,{\rm arc min}$ at 1~GHz), an under-illuminated primary reflector of about 40~m projected aperture diameter, D, is required\footnote{The equivalent fully illuminated aperture would be $D\approx 25\,{\rm m}$}. If the telescope is to have no moving parts, i.e. fixed pointing relative to the ground, then a wide instantaneous field-of-view (FOV) with multiple feeds is essential so that as the Earth rotates a broad strip of sky can be mapped during the course of a day. To reach the target sensitivity, a minimum number of 50 pixels is required in the focal plane, each of them reaching the following performance:
\begin{itemize}
\item Forward gain loss in comparison to central pixel  $<1\,{\rm dB}$;
\item Cross-polarisation better than $-30\,{\rm dB}$;
\item Beam ellipticity lower than about 0.1.
\end{itemize}

We propose to adopt an offset parabolic design in order to avoid blockage, minimise diffraction and scattering from the struts, and also to avoid standing waves that might be created between the dish and the feeds. A wide FOV  with the required optical performance mentioned above can only be achieved with a long focal length. We found that a focal length $F=90\,{\rm m}$, giving a focal ratio $F/D=2.3$, much larger than more conventional radio telescopes (having typically a $F/D$ ratio of about 0.4), is suitable for our purpose. This has the drawback of requiring the use of feeds with a small full-width-half-maximum, therefore with a large aperture diameter. For our telescope design, this means feeds with an aperture diameter of at least $2\,{\rm m}$ producing an edge taper telescope illumination of $-20\,{\rm dB}$ in order to reduce the spillover. We are targeting a spillover of less than 2\% in total power as the fraction of the horn beam not intercepting the telescope will see radiation from the ground therefore increasing the system temperature $T_{\rm sys}$.

\begin{figure}
\begin{center}
 \includegraphics[angle=0,scale=0.5]{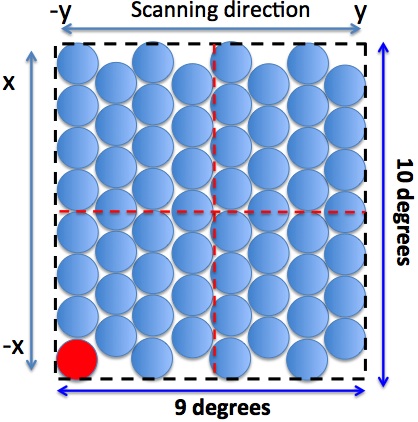}
\caption{Focal plane of staggered feeds. A total of 60 pixels laid out in a rectangle (dashed black line) of $16\,{\rm m}\times 15\,{\rm m}$. This will allow for a FOV of about $10\times 9\,{\rm deg}^2$. Axes $x$ and $y$ are shown as a reference to the positions mentioned in Table~\ref{opt_perf}. The pixel shown in red is the off-axis pixel corresponding to the beam profiles in Fig.~\ref{Beams}.}
\label{FPU}
\end{center}
\end{figure}

The design adopted for the focal plane is given in Fig.~\ref{FPU}. A line of eight feeds allows for coverage of a $10\,{\rm deg}$ strip. Along the scanning direction produced by the Earth's rotation, other lines of feeds are staggered to allow for a proper sampling on the sky. With this configuration, 60 pixels, 10 more than the requirement, will produce a $10\times 9~{\rm deg}^2$ FOV. Assuming a feed aperture diameter of about $2\,{\rm m}$, the resulting focal plane will be comprised within a rectangle of  $16~{\rm m}\times 15~{\rm m}$.

\begin{figure}
\begin{center}
 \includegraphics[angle=0,scale=0.25]{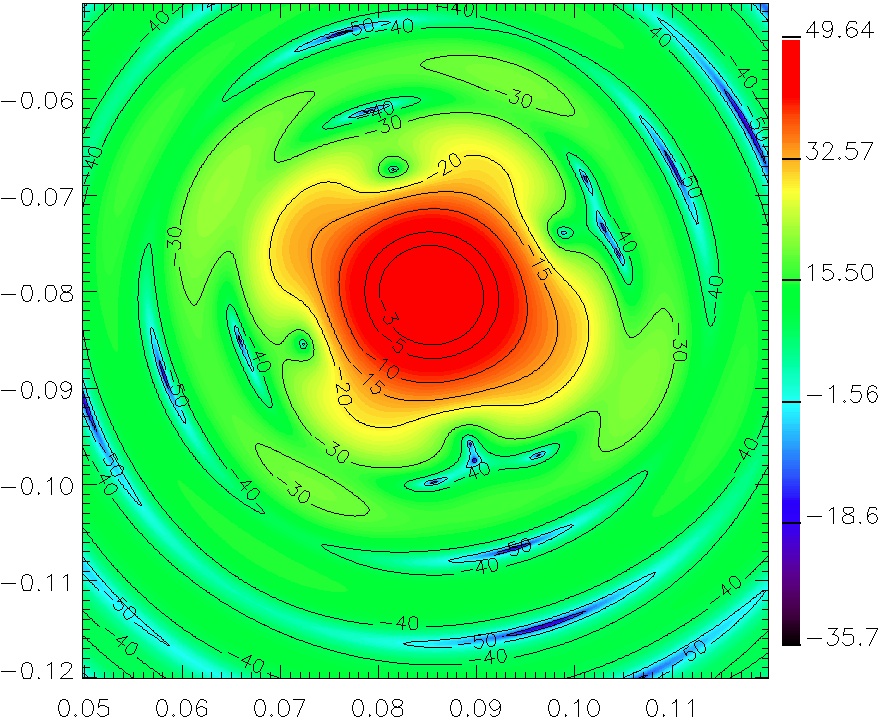}
  \includegraphics[angle=0,scale=0.25]{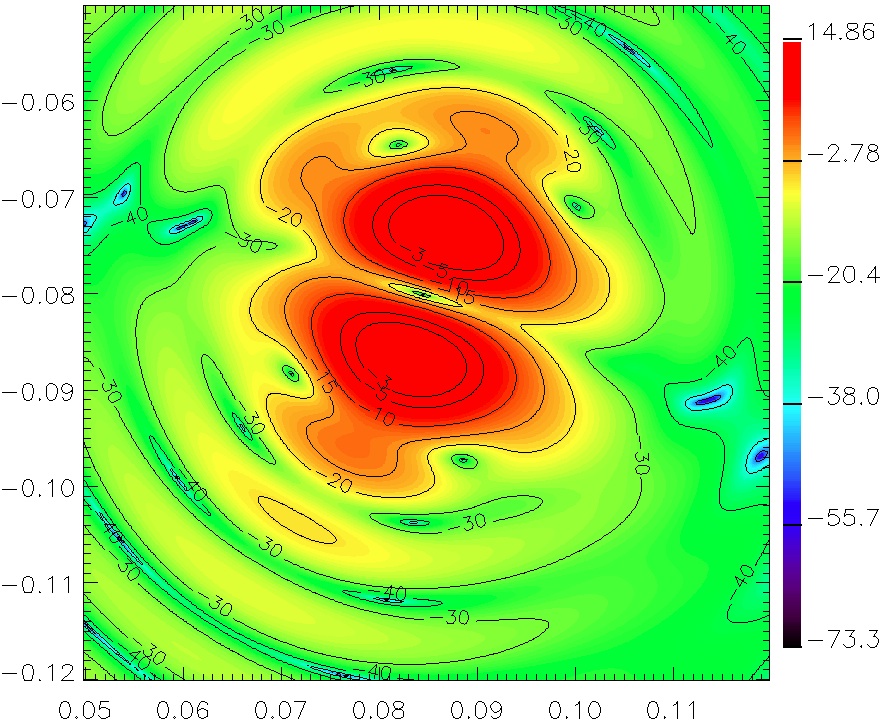}
\caption{U-V maps of beam intensity for the most off-axis pixel corresponding to the focal plane position $x=-8{\rm m}$ and $y=-7.5\,{\rm m}$ (represented in red in Fig~\ref{FPU}). Top: co-polarisation; bottom: cross-polarisation. Simulations performed using ideal telescope and horn. Colour scale:  the gain in dB relative to maximum.}
\label{Beams}
\end{center}
\end{figure}

Optical simulations of such a concept have been performed using Physical Optics (GRASP from TICRA) in order to take into account the diffraction effects. Results from this model, assuming ideal components (perfect telescope and Gaussian feedhorns) and a perfect alignment, are given in Table~\ref{opt_perf}. The performance is well within the requirements even for the most off-axis pixel (represented in red in Fig.~\ref{FPU}).
 Fig.~\ref{Beams} represents the modelled contour plot of the beams (co- and cross-polarisation beams) in the U-V plane that can be expected for the pixel located at the corner edge of the focal plane. Taking into account degradations due to using real components and with typical alignment errors, we can till expect to meet the requirements.

\begin{table}
\caption{Optical performance calculated using GRASP assuming ideal telescope and Gaussian feeds with -20~dB edge taper. Results are given for a pixel that would be located at the centre of the focal plane and for the corner edge pixel (red in Fig.~\ref{FPU} at $x=-8\,{\rm m}; y=-7.5\,{\rm m}$). Peak cross-polarisation is given respectively to the co-polarisation forward gain.}
\begin{center}
\begin{tabular}{ccccc}
\hline
Pixel &Forward&Ellipticity&Peak&FWHM\\
Position & Gain &  & X-Pol & \\
\hline
&dB&&dB&arcmin\\
\hline
Centre&49.8&10$^{-3}$&-40&38\\
Edge&49.6& 0.07&-35&39\\ 
\hline
\end{tabular}
\end{center}
\label{opt_perf}
\end{table}

\begin{figure*}
\includegraphics[width=0.5\textwidth]{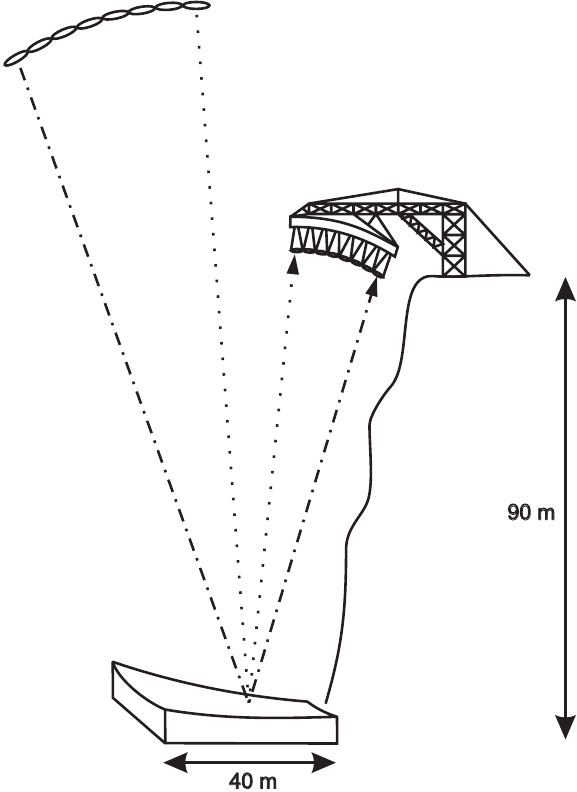}
\caption{Schematic of the proposed design of the BINGO telescope. There will be an under-illuminated $\sim 40\,{\rm m}$ static parabolic reflector at the bottom of a cliff which is around $\sim  90\,{\rm m}$ high. At the top of the cliff will be placed a boom on which the receiver system of $60$ feed-horns will be mounted.}
\label{telescope}
\end{figure*}

The design concept is illustrated in Fig.\ref{telescope} and the key parameters listed in Table~\ref{tab:telescope}. Rather than a high tower to support the focal arrangement, the plan would be to identify a suitable site in which the dish structure can be built below a cliff (or near vertical slope) of height $90\,{\rm m}$ so that the feed arrangement can be fixed on a boom near the top of the cliff. 

\begin{table}
\caption{Summary of proposed BINGO telescope parameters.}
\begin{center}
\begin{tabular}{cc}
\hline
Reflector diameter&40~m\\
Resolution (at $\lambda=30\,{\rm cm}$)&40 arcmin\\
Number of feeds&  60\\
Instantaneous field of view&10~deg$\times$9~deg\\
Frequency range&960 MHz to 1260 MHz\\
Number of frequency channels&$\geq$ 300\\
\hline

\end{tabular}
\end{center}
\label{tab:telescope}
\end{table}%

\subsection{Design for a receiver module}

\noindent A high degree of stability is an essential design criterion
both for telescope and receivers. Such demands are not unique since
instruments used to measure the structure in the CMB have been built
which meet similar stability criteria. In CMB experiments correlation
receivers (or interferometers) are generally the design of choice
(for example, {\it WMAP} and {\it Planck}).  They can reduce the effect of gain
fluctuations in the amplifiers ($1/f$ noise) by up to three orders of
magnitude \citep{Mennella:2010}.

\begin{figure*}
\includegraphics[width=0.8\textwidth]{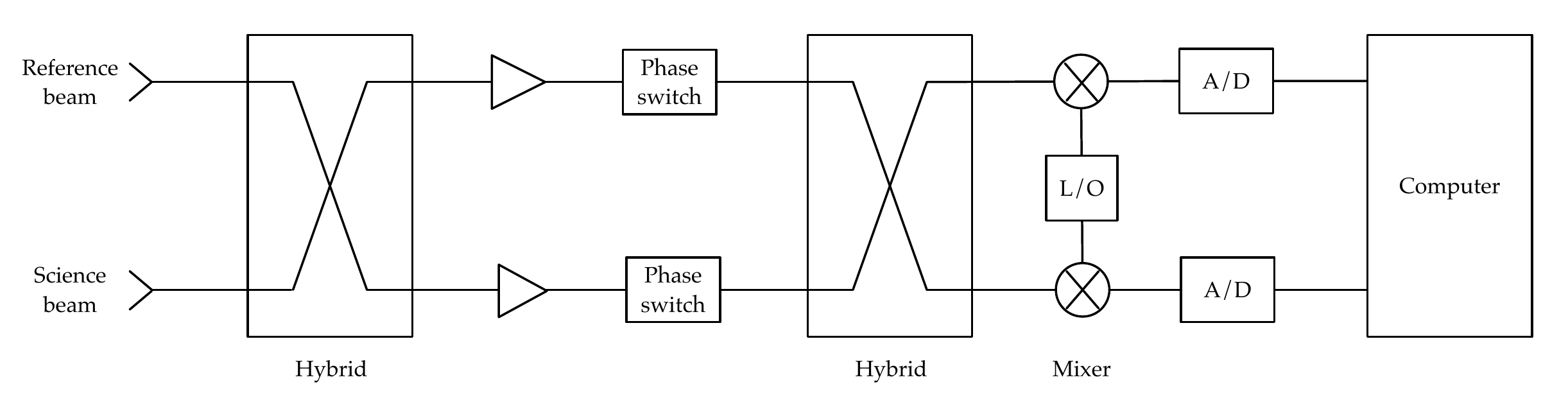}
\caption{Block diagram of the receiver chain. The reference beam will point towards one of the 
Celestial Poles. The hybrids will be waveguide magic tees. After the
second hybrid the signal will be down-converted using mixers with a
common local oscillator. There will be further amplification before digitization. }
\label{receiver}
\end{figure*}

We plan to adopt the same approach with each receiver module producing
a ``difference spectrum'' between the desired region of sky and a
reference signal.  The choice of the reference is important. Ideally
it should have the same spectrum as the sky and be close to the same
brightness temperature. It must also be very stable. An obvious choice
is another part of the sky. We propose that there should be a
reference antenna for each module with a beamwidth of around
$20^{\circ}$ which would have a beam area sufficiently large to
average out all but the lowest $\ell$ modes (see Fig. \ref{Fig:HIspec}). It should be fixed and we suggest that it should
point directly to one of the celestial poles. In this way all
measurements at all times of day, and all the year round, would then
be compared to the same region of sky. Since the sky emission has
significant linear polarization ($\sim 10\,\%$ or more;
\citet{Wolleben:2006}), measuring with circularly polarized feeds  is
necessary. Similar considerations will be important for foreground
emission removal.  A schematic diagram for a single
polarization channel of the proposed receiver module design is shown
in Fig.~\ref{receiver}.

\subsection{Receiver simulations}

\begin{figure*}
\centering
\subfigure[]{\label{fig:white}\includegraphics[width=0.45\textwidth]{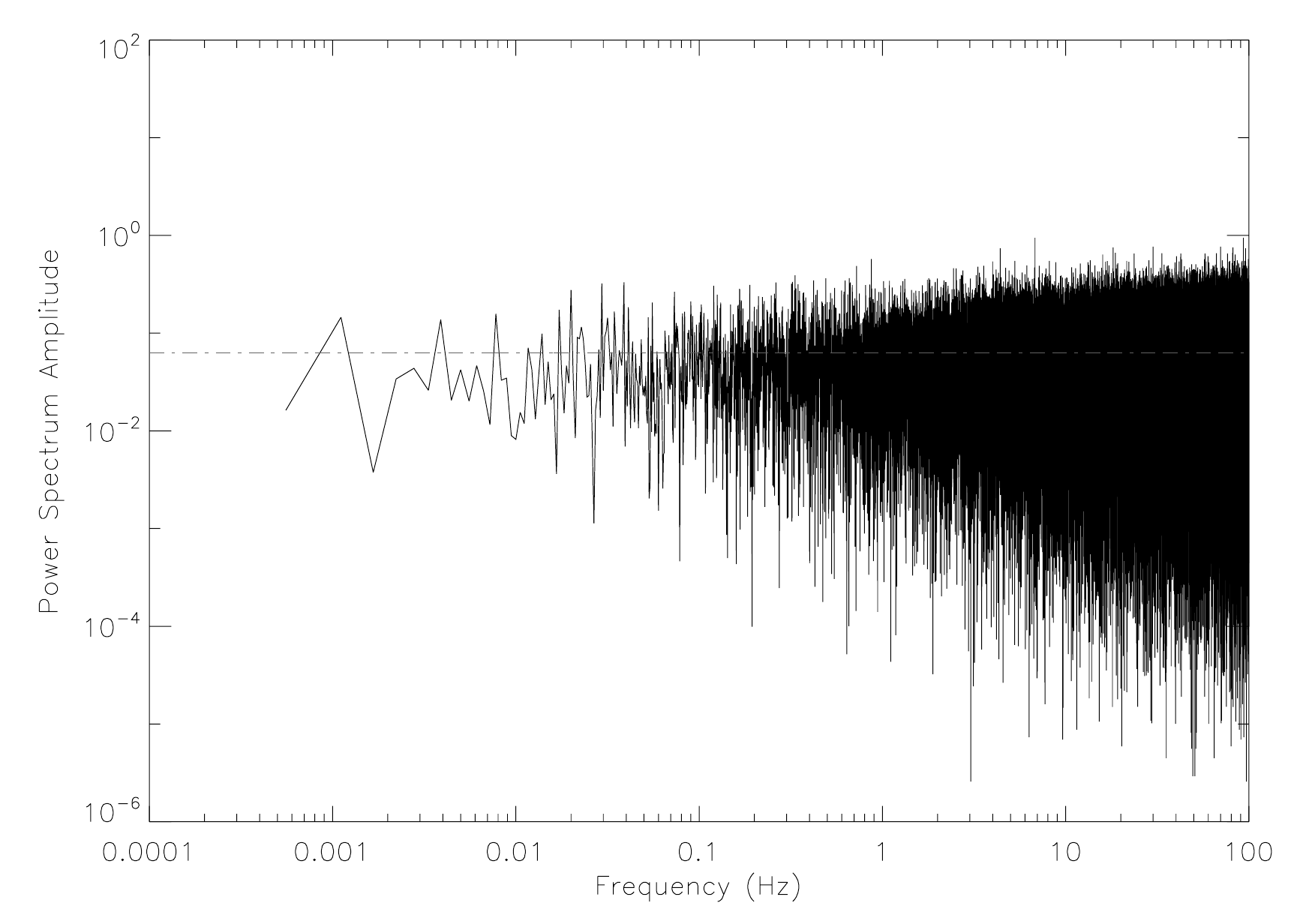}}
\subfigure[]{\label{fig:pink}\includegraphics[width=0.45\textwidth]{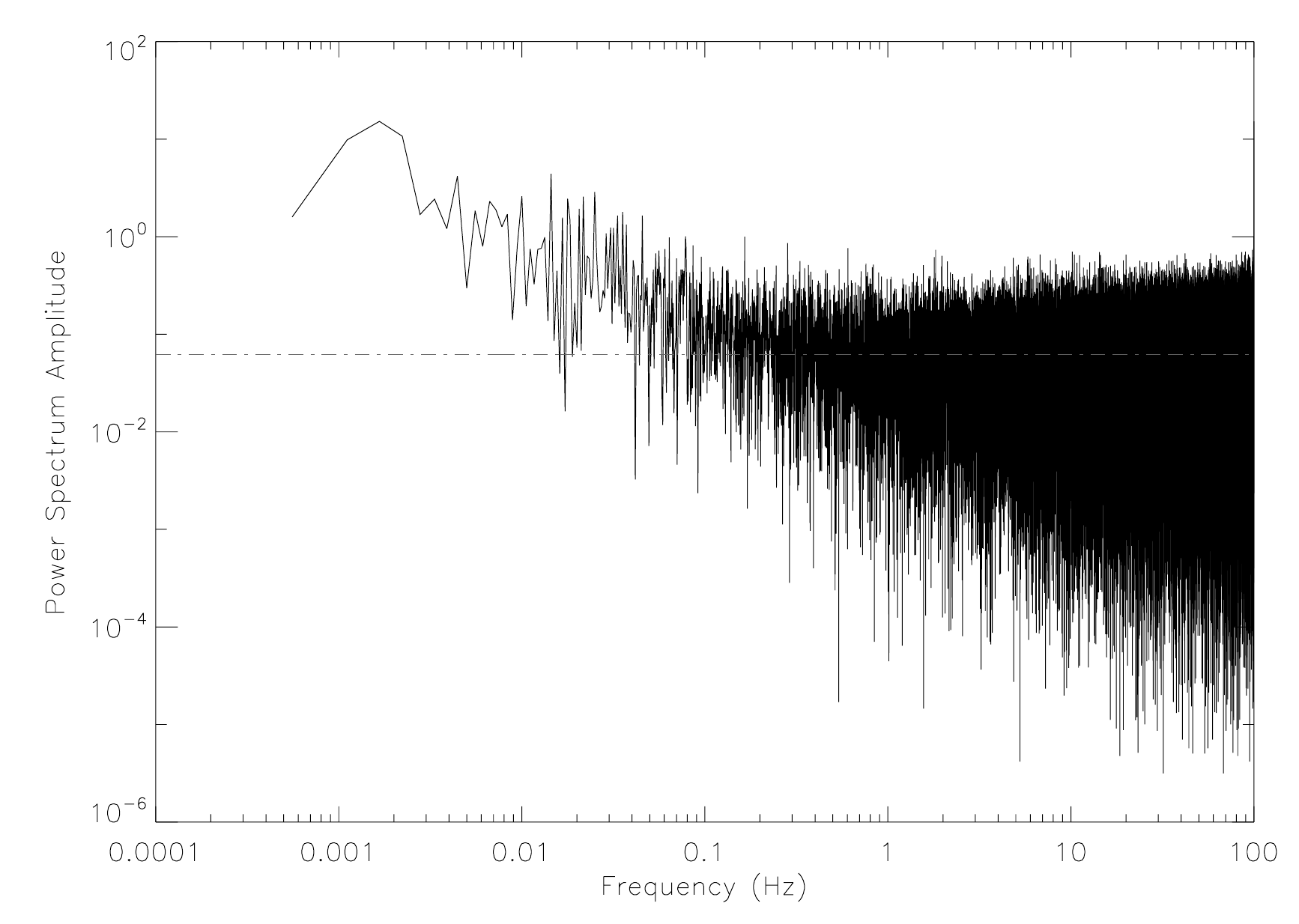}}
\caption{Simulated noise power spectra for a realistic BINGO receiver which uses amplifiers with $f_{\rm knee}=1\,{\rm Hz}$ and a $30\,{\rm min}$ integration. Here (a) represents a perfectly balanced receiver; and (b) a receiver with imbalances in both system temperature, $(A_{\mathrm sys}-B_{\mathrm sys})/A_{\mathrm sys}=0.2$  and power splitting, $\epsilon=0.4$. These values are significantly larger than expected in a realistic situation and have been chosen to illustrate the point. The upturn of the spectrum in (b) indicates the presence of $1/f$ noise. }
\label{perfvimperf}
\end{figure*}

An ideal correlation receiver - one with perfectly matched system
temperatures and no imbalance in the power splitting within each
hybrid - can effectively remove all $1/f$ noise from its signal
output. If $A=A_{\rm sys}+\delta A$ and $B=B_{\rm sys}+\delta B$ are
the total voltages entering the receiver chain and $A_{\rm sys}\propto\sqrt{T_{\mathrm sys}^A}$ and
$B_{\rm sys}\propto\sqrt{T_{\mathrm sys}^B}$ are the system temperature contributions and $\delta A$
and $\delta B$ are the signals we want to detect, then the difference
detected power at the end of the chain is proportional to the
difference, $\delta A^{2} - \delta B^{2}$, between the science ($A$) and
reference beam ($B$). The spectrum of such a signal
exhibits no excess power at low frequencies ($1/f$ noise) as illustrated
in Fig.~\ref{fig:white}, and integration can continue indefinitely
with the white noise falling as the square root of integration time.

However, real life receivers are imperfect. Imbalances between system
temperature and power splitting both limit the reduction of the $1/f$
noise resulting in a break in the power spectrum quantified in terms of
a knee frequency \citep{Seiffert:2002} separating white
noise-dominated and pink noise-dominated parts of the spectrum.
Consequently there will be a limit to the viable integration time as
illustrated in Fig.~\ref{fig:pink}. It is useful to quantify by how
much a correlation receiver is limited by these imperfections, and to
do so we simulate their effect in a realistic receiver model. We
assess the maximum permissible imperfection amplitude to ensure the
knee frequency remains below $f_\mathrm{knee}\sim1$ mHz, a value
corresponding $\sim 20$~minutes integration time - sufficient for the
largest HI structures we are interested in detecting, $\sim \hbox{a
few degrees}$, to drift through the telescope beam.

We include Gaussian (white) and $1/f$ noise, generating the former
using a simple random number procedure and the latter with a more
complex {\it auto-regressive} approach - essentially filtering a white
noise time stream into one with correct correlation properties.  The
knee frequency of component amplifiers depends on both the physical
choice of amplifier and the channel bandwidth. Though the actual value
will need to be determined empirically, we will assume in the current
simulations that the individual amplifier knee frequencies are
$f_\mathrm{knee}^\mathrm{amp}\sim1\,{\rm Hz}$ for a channel bandwidth of
$1\,{\rm MHz}$.

\subsubsection{Unbalanced $T_\mathrm{sys}$}

In an ideal receiver the two
channels, $A_\mathrm{sys}$ and $B_\mathrm{sys}$, should be
perfectly matched. This is impossible to achieve at all times due to the
angular variations in the Galactic emission and here we investigate
the effect on the knee frequency due to small imbalances between the
temperatures.

Modelling the entire correlation receiver and simulating noise spectra
for the output difference signal in the presence of small temperature
imbalances, we find that the knee frequency of the correlation
receiver can be approximated by $f_\mathrm{knee}=1.2\,{\rm
Hz}\,((A_{\mathrm sys}-B_{\mathrm sys})/A_{\mathrm sys})^2$ and therefore to
ensure BINGO's knee frequency remains continually below the desired
$f_\mathrm{knee}\sim1\,{\rm mHz}$, our simulations show that $(A_{\mathrm
sys}-B_{\mathrm sys})/A_{\mathrm sys}$ needs to be $< 0.03$. In practice
the balance can be maintained electronically so this should not be an
important limitation, provided the unbalance behaves predictably as a
function of time. The main source of imbalance in the system
temperatures will arise from the changing sky background as the
telescope scans across the sky. Typical variations in those parts of
the sky of interest well away from the Galactic plane will be a few
Kelvin and will be repeatable from day to day. Importantly the sky
emission will be spectrally smooth and virtually identical in all
frequency channels. This means that once the frequency-averaged
variation of sky brightness has been measured over the course of a
siderial day, a multiplicative correction, the $r$-factor defined by
\citep{Seiffert:2002}, can be applied to all channels to bring the $A$
and $B$ outputs into balance before differencing, thus ensuring the
$(A_{\mathrm sys}-B_{\mathrm sys})/A_{\mathrm sys}$ $< 0.03$ is met.

\subsubsection{Power Split Imbalance}

We now consider the case of hybrids in which the power
splitting is imperfect. To do so, we introduce the constant
coefficient $\epsilon$, representing an imbalance
between the power emerging from the two output ports of the
hybrids. The signals emerging from such a hybrid are then
\begin{eqnarray}
A_\mathrm{out}&=&\sqrt{\tfrac{1-\epsilon}{2}}A+\sqrt{\tfrac{1+\epsilon}{2}}B~,\nonumber \\
B_\mathrm{out}&=&\sqrt{\tfrac{1+\epsilon}{2}}A-\sqrt{\tfrac{1-\epsilon}{2}}B~,
\end{eqnarray}
\\where $A,B$ are the two input signals.

Here we assume the power splitting imbalances in the first and second
hybrids are equal, and simulate noise spectra for different values of $\epsilon$. Fitting for the knee frequency we find
$f_\mathrm{knee}=4\epsilon^2\,{\rm Hz}$. Again, in order to maintain a
knee frequency of 1 mHz, this relationship implies that the power
splitting imbalance should satisfy $\epsilon< 0.015$. This requirement
can be relaxed if a $180^\circ$ phase switch is used within the hybrid loop and,
since waveguide magic-tees typically have a power split imbalance of
$\epsilon= 0.02$, the required performance should be achievable.

\subsection{Practicalities}

\noindent Our philosophy is to choose simplicity while keeping in mind
good engineering practice and, wherever possible, to adopt
conventional solutions to problems rather than try to invent new
ones. 

To reduce the impact of RFI the telescope would need to be in a
remote site and therefore an important additional consideration is that
everything should be easy to maintain. The design, construction and
mounting of the feeds is perhaps the most challenging aspect of the
overall concept. As stated above, low sidelobe levels are important in
order to mitigate the effects of RFI (even in the remotest part of the
World) and ease the separation of foreground emission from the desired
neutral hydrogen signal. If a conventional
horn design is adopted, these horns will have to be extremely large;
to illuminate a $30\,{\rm m}$ dish from a focus $\sim 75\,{\rm m}$ away will require
an aperture diameter for each feed of 1.5 to 2~m and a length of
$\sim 8\,{\rm m}$. Such feeds are practical, in the sense that they would be
scaled up versions of horns built to existing feed designs, but could
prove expensive to build and unwieldy. Alternatives like using small
secondary mirrors for each receiver module are under consideration.

An integral part of the feed system is the extraction of polarized
signals which we will do by extracting the two opposite circular polarizations.
Minimizing any Ohmic loss before the first amplifiers is
essential if good thermal noise performance is to be achieved. After
the feed the next element is the hybrid. Due to their low loss and
good power splitting properties we would use waveguide magic-tees.

We would use non-cryogenic low noise amplifiers because at this
frequency the noise advantages of cooling to $\sim 20\,{\rm K}$ are outweighed
by the low cost, the low power consumption and the low maintenance
demands of a non-cryogenic approach. It would also be desirable to
integrate the amplifiers with the magic-tee to avoid the necessity to
use potentially lossy cables. Some cooling, perhaps using Peltier
coolers, could be used because it would provide both
temperature stabilization for the amplifiers and a potential
improvement in noise performance.

An important choice is when to digitize the signal. In conventional
correlation receivers the ``correlation'' is performed in a second
hybrid. Our preferred  solution is to down-convert the radio frequency
signal with a common local oscillator and then to digitize and perform
the function of the second hybrid in software using FPGAs to multiply
the signals and produce averaged spectra which can then be
differenced. One advantage of such a system is that the levels in the
two arms of the receiver can be easily adjusted so that they are the
same and this then operates the receiver at its optimum for gain
fluctuation reduction.

\section{Conclusions}

Intensity mapping is  an innovative idea which has been suggested to allow large-scale structure surveys using the 21cm line of neutral hydrogen using instruments with relatively small collecting area compared to the SKA. Such concepts would naturally be much cheaper. We have presented calculations that illustrate the characteristics of the signal and it is shown that it should have an r.m.s. of $\sim 100\,\mu{\rm K}$ on angular scales $\sim 1\,{\rm deg}$ and bandwidths of $1-10\,{\rm MHz}$. At  very low redshifts there will be a shot noise effect due to it being dominated by individual galaxies, at medium redshifts it is dominated by the effects of HI in galaxy clusters and at high redshifts becomes an unresolved background. 

We have performed an optimisation analysis which suggests that a significant measurement of the acoustic scale can be made between $z=0.13-0.48$ using a single dish telescope of diameter $40\,{\rm m}$ with a drift scan survey of a strip with width $10\,{\rm deg}$ covering a total of $\sim 2000\,{\rm deg}^2$.  In order to do this at higher redshift (specifically, the case of $z=0.58-1.47$) would require a much larger dish, $\sim 140\,{\rm m}$, and an area of $500\,{\rm deg}^2$ which would probably be best done using an interferometer in order to allow non-drift scan observations using such a long baseline instrument. We have presented a detailed assessment of the likely foregrounds and conclude that there will be continuum fluctuations with an r.m.s of $\sim 100\,{\rm mK}$ on degree scale, but that this should not significantly affect our ability to extract the HI signal due to the fact that its spectral profile is expected to be relatively smooth.

We have put forward a design concept, which we have named BINGO, to perform the lower of the two redshift surveys studied. This will use a large focal length telescope, $F/D\approx 2.3$, to mount  an array of $60$ feedhorns approximately $90\,{\rm m}$ above a static $40\,{\rm m}$. It will use pseudo-correlation receivers to reduce $1/f$ noise due to gain fluctuations. We are currently searching for a site and the funding needed to bring this idea to fruition.

\section*{Acknowledgements}
We thank Peter Dewdney for suggesting the use of a cliff to mount the
BINGO feeds  and Xiang-Ping Wu for suggesting using one of the 
Celestial Poles as a reference. We have had numerous useful discussions
with Peter Wilkinson, Shude Mao and Richard Davis. CD acknowledges an
STFC Advanced Fellowship and an EC IRG grant under the FP7. AP is supported by the project GLENCO, funded under the
FP7, Ideas, Grant Agreement n. 259349 and also received
support from the STFC(UK). We thank Bob Watson for help in creating Fig.~\ref{telescope}.

\bibliographystyle{natbib}

\end{document}